%%%%%%%%%%%%%%%%%%%%%%%%%%%%%%%%%%%%%%%%%%%%%%%%%%%%%%%%%%%%%%%%%%%%%%%%%
%                                                                        % 
%        by J. Bouttier, P. Di Francesco and  E. Guitter                 %
%                TEX file, using lanlmac.tex macros                      %
%                                                                        %
%                                                                        %
%%%%%%%%%%%%%%%%%%%%%%%%%%%%%%%%%%%%%%%%%%%%%%%%%%%%%%%%%%%%%%%%%%%%%%%%%%
\input lanlmac
\def\href#1#2{{#2}}

\input epsf.tex

\overfullrule=0mm

\newcount\figno
\figno=0
\def\fig#1#2#3{
\par\begingroup\parindent=0pt\leftskip=1cm\rightskip=1cm\parindent=0pt
\baselineskip=11pt
\global\advance\figno by 1
\midinsert
\epsfxsize=#3
\centerline{\epsfbox{#2}}
\vskip 12pt
{\bf Fig.\ \the\figno:} #1\par
\endinsert\endgroup\par
}
\def\figlabel#1{\xdef#1{\the\figno}}
\def\encadremath#1{\vbox{\hrule\hbox{\vrule\kern8pt\vbox{\kern8pt
\hbox{$\displaystyle #1$}\kern8pt}
\kern8pt\vrule}\hrule}}

%Macros 
%%%%%%%%%%%%%%%%%%%%%%%%%%%%%%%%%%%%%%%%%%%%%%%%%%%%%%%%%%%%%%%%%

\def\IR{\relax{\rm I\kern-.18em R}}
\font\cmss=cmss10 \font\cmsss=cmss10 at 7pt

\font\cmss=cmss10 \font\cmsss=cmss10 at 7pt
\def\IZ{\relax\ifmmode\mathchoice
{\hbox{\cmss Z\kern-.4em Z}}{\hbox{\cmss Z\kern-.4em Z}}
{\lower.9pt\hbox{\cmsss Z\kern-.4em Z}}
{\lower1.2pt\hbox{\cmsss Z\kern-.4em Z}}\else{\cmss Z\kern-.4em Z}\fi}
\def\IN{\relax{\rm I\kern-.18em N}}
\def\circbullet{{\bigcirc \kern-.75em \bullet \kern .3em}}
\def\circcirc{{\bigcirc \kern-.75em \circ \kern.3em}}
\def\smallcircbullet{{{\scriptscriptstyle{\bigcirc \kern-.5em \bullet}} \kern .2em}}
\def\smallcirccirc{\relax{{\scriptscriptstyle{\bigcirc \kern-.5em \circ}} \kern .2em}}
\def\b{\circ}
\def\n{\bullet}

\def\gbbbb{\Gamma_4^{\hbox{$\scriptstyle \b \b$}\kern -8.2pt
\raise -4pt \hbox{$\scriptstyle \b \b$}}}
\def\gnnnn{\Gamma_4^{\hbox{$\scriptstyle \n \n$}\kern -8.2pt  
\raise -4pt \hbox{$\scriptstyle \n \n$}}}
\def\gnnnnnn{\Gamma_6^{\hbox{$\scriptstyle \n \n \n$}\kern -12.3pt
\raise -4pt \hbox{$\scriptstyle \n \n \n$}}}
\def\gbbbbbb{\Gamma_6^{\hbox{$\scriptstyle \b \b \b$}\kern -12.3pt
\raise -4pt \hbox{$\scriptstyle \b \b \b$}}}
\def\gbbbbc{\Gamma_{4\, c}^{\hbox{$\scriptstyle \b \b$}\kern -8.2pt
\raise -4pt \hbox{$\scriptstyle \b \b$}}}
\def\gnnnnc{\Gamma_{4\, c}^{\hbox{$\scriptstyle \n \n$}\kern -8.2pt
\raise -4pt \hbox{$\scriptstyle \n \n$}}}
\def\Rbud#1{{\cal R}_{#1}^{-\kern-1.5pt\blacktriangleright}}
\def\Rleaf#1{{\cal R}_{#1}^{-\kern-1.5pt\vartriangleright}}
\def\Rbudb#1{{\cal R}_{#1}^{\circ\kern-1.5pt-\kern-1.5pt\blacktriangleright}}
\def\Rleafb#1{{\cal R}_{#1}^{\circ\kern-1.5pt-\kern-1.5pt\vartriangleright}}
\def\Rbudn#1{{\cal R}_{#1}^{\bullet\kern-1.5pt-\kern-1.5pt\blacktriangleright}}
\def\Rleafn#1{{\cal R}_{#1}^{\bullet\kern-1.5pt-\kern-1.5pt\vartriangleright}}
\def\Wleaf#1{{\cal W}_{#1}^{-\kern-1.5pt\vartriangleright}}
\def\Cleaf{{\cal C}^{-\kern-1.5pt\vartriangleright}}
\def\Cbud{{\cal C}^{-\kern-1.5pt\blacktriangleright}}
\def\Crleaf{{\cal C}^{-\kern-1.5pt\circledR}}

%%%%%%%%%%%%%%%%%%%%%%%%%%%%%%%%%%%%%%%%%%%%%%%%%%%%%%%%%%%%%%%%%
%%%%%%%%%%%%%%%%%%%%%%%%%%%%%%%%%%%%%%%%%%%%%%%%%%%%%%%%%%%%%%%%%%%%%

\magnification=\magstep1
\baselineskip=12pt
\hsize=6.3truein
\vsize=8.7truein
 at 8truept
 at 8truept
 at 10truept

%%%%%%%%%%%%%%%%%%%%%%%%%%%%%%%%%%%%%%%%%%%%%%%%%%%%%%%%%%%%%%%%%%%%%%%%
\font\bigrm=cmr12 at 14pt
\centerline{\bigrm Multicritical continuous random trees}

\bigskip\bigskip

\centerline{J. Bouttier${}^{1,2}$, P. Di Francesco${}^1$ and E. Guitter${}^1$}
  \medskip
  \centerline{\it ${}^1$Service de Physique Th\'eorique, CEA/DSM/SPhT}
  \centerline{\it Unit\'e de recherche associ\'ee au CNRS}
  \centerline{\it CEA/Saclay}
  \centerline{\it 91191 Gif sur Yvette Cedex, France}
  \smallskip
  \centerline{\it ${}^2$Instituut voor Theoretische Fysica}
  \centerline{\it Valckeniertstraat 65}
  \centerline{\it 1018 XE Amsterdam}
  \centerline{\it The Netherlands}
  \medskip
\centerline{\tt bouttier@spht.saclay.cea.fr}
\centerline{\tt philippe@spht.saclay.cea.fr}
\centerline{\tt guitter@spht.saclay.cea.fr}

  \bigskip
  \bigskip
  \bigskip

%  \centerline{\footrm Submitted: Jan 1, 2004;  Accepted: Jan 2, 2004;
%     Published: Jan 3, 2004}
%     \centerline{\footrm Mathematics Subject Classifications: Primary 05C30; Secondary 05A15,
%     05C05, 05C12, 68R05}

     \bigskip\bigskip

     \centerline{\bf Abstract}
     \smallskip
     {\narrower\noindent
We introduce generalizations of Aldous' Brownian Continuous Random Tree as scaling limits
for multicritical models of discrete trees. These discrete models involve trees with fine-tuned
vertex-dependent weights ensuring a $k$-th root singularity in their generating function.
The scaling limit involves continuous trees with branching points of order up to $k+1$. 
We derive explicit integral representations for the average profile of this $k$-th order
multicritical continuous random tree, as well as for its history distributions measuring multi-point 
correlations. The latter distributions involve non-positive universal weights at the branching points
together with fractional derivative couplings. We prove universality by rederiving the same results 
within a purely continuous axiomatic approach based on the resolution of a set of consistency 
relations for the multi-point correlations. The average profile is shown to obey a fractional 
differential equation whose solution involves hypergeometric functions and matches the integral 
formula of the discrete approach.
\par}

   \bigskip
   \bigskip
   \bigskip

%\draft
%AMS Subject Classification (2000): Primary 05C30; Secondary 05A15, 
%05C05, 05C12, 68R05

%references
\nref\SCH{G. Schaeffer, {\it Bijective census and random
generation of Eulerian planar maps}, Elec.
Jour. of Combinatorics Vol. {\bf 4} (1997) R20.}
\nref\CONST{M. Bousquet-M\'elou and G. Schaeffer,
{\it Enumeration of planar constellations}, Adv. in Applied Math.,
{\bf 24} (2000) 337-368.}
\nref\COL{J. Bouttier, P. Di Francesco and E. Guitter, {\it Counting colored random
triangulations}, Nucl. Phys. {\bf B641[FS]} (2002) 519-532, arXiv:cond-mat/0206452.}
\nref\CENSUS{J. Bouttier, P. Di Francesco and E. Guitter, {\it Census of planar
maps: from the one-matrix model solution to a combinatorial proof},
Nucl. Phys. {\bf B645}[PM] (2002) 477-499, arXiv:cond-mat/0207682.}
\nref\CS{P. Chassaing and G. Schaeffer, {\it Random Planar Lattices and 
Integrated SuperBrownian Excursion}, 
Probability Theory and Related Fields {\bf 128(2)} (2004) 161-212, 
arXiv:math.CO/0205226.}
\nref\MOB{J. Bouttier, P. Di Francesco and E. Guitter, {\it Planar
maps as labeled mobiles}, Elec. Jour. of Combinatorics Vol. {\bf 11} (2004) R69,
arXiv:math.CO/0405099.}
\nref\GEOD{J. Bouttier, P. Di Francesco and E. Guitter, {\it Geodesic
distance in planar graphs}, Nucl. Phys. {\bf B663}[FS] (2003) 535-567, 
arXiv:cond-mat/0303272.}
\nref\QGRA{V. Kazakov, {\it Bilocal regularization of models of random
surfaces}, Phys. Lett. {\bf B150} (1985) 282-284; F. David, {\it Planar
diagrams, two-dimensional lattice gravity and surface models},
Nucl. Phys. {\bf B257} (1985) 45-58; J. Ambjorn, B. Durhuus and J. Fr\"ohlich,
{\it Diseases of triangulated random surface models and possible cures},
Nucl. Phys. {\bf B257}(1985) 433-449; V. Kazakov, I. Kostov and A. Migdal
{\it Critical properties of randomly triangulated planar random surfaces},
Phys. Lett. {\bf B157} (1985) 295-300.}
\nref\BIPZ{E. Br\'ezin, C. Itzykson, G. Parisi and J.-B. Zuber, {\it Planar
Diagrams}, Comm. Math. Phys. {\bf 59} (1978) 35-51.}
\nref\DGZ{P. Di Francesco, P. Ginsparg and J. Zinn--Justin, {\it 2D Gravity and Random Matrices},
Physics Reports {\bf 254} (1995) 1-131.}
\nref\BS{for a review, see J.F. Le Gall {\it Spatial Branching Processes, Random Snakes
and Partial Differential Equations} Birkh\"auser (1999)}
\nref\ISE{D.J. Aldous, {\it Tree-Based Models for Random Distribution of Mass}, J.Stat.Phys.
{\bf 73} (1009), no. 3/4,  625-641.}
\nref\AL{J. Ambj\o rn and R. Loll, {\it Non-perturbative Lorentzian
Quantum Gravity, Causality and Topology Change},
Nucl. Phys. {\bf B536 [FS]} (1998) 407, hep-th/9805108.}
\nref\LORGRA{P. Di Francesco and E. Guitter, {\it Critical and Multicritical Semi-Random $(1+d)$-
Dimensional Lattices and Hard Objects in $d$ Dimensions}, J.Phys. A35 (2002) 897-928.}
\nref\ALDOUS{D.J. Aldous, 
{\it The continuum random tree. I.} Ann. Probab. 19 (1991), no. 1, 1--28;
{\it The continuum random tree. II. An overview.} in Stochastic analysis 
(Durham, 1990),  23--70, London Math. Soc. Lecture Note Ser., 167, Cambridge Univ. Press, Cambridge, 1991;
{\it The continuum random tree. III.} Ann. Probab. 21 (1993), no. 1, 248--289.}
\nref\BMS{M. Bousquet-M\'elou and G. Schaeffer,{\it The degree distribution
in bipartite planar maps: application to the Ising model},
arXiv:math.CO/0211070.}
\nref\CHP{J. Bouttier, P. Di Francesco and E. Guitter, {\it Combinatorics of 
hard particles on planar graphs}, Nucl. Phys. {\bf B655}[FS] (2003) 313-341, 
arXiv:cond-mat/0211168.}
\nref\HARGRA{J. Bouttier, P. Di Francesco and E. Guitter, {\it  Combinatorics of bicubic maps with 
hard particles}, J.Phys. A38 (2005) 4529-4560.}
\nref\ONEMATH{E. Br\'ezin and V. Kazakov, {\it Exactly solvable field theories of closed strings}, 
Phys. Lett. B236 (1990) 144-150; M. Douglas and S. Shenker, {\it Strings in less than 1 dimension}, 
Nucl. Phys. B335 (1990) 635; D. Gross and A. Migdal, {\it Non-perturbative two-dimensional gravity}, 
Phys. Rev. Lett. 64 (1990) 127-130.}
\nref\HILF{see for instance P.L. Butzer and U. Westphal, {\it An introduction to fractional calculus}
in {\it Applications of fractional calculus in physics}, R. Hilfer editor, Word Scientific (2000)
1-85.}
\nref\WOLF{see for instance http://functions.wolfram.com/HypergeometricFunctions/Hyper\-geo\-metricPFQ/06/02/01/ }
\nref\CARDY{J. Cardy, {\it Conformal invariance and the Yang-Lee edge
singularity in two dimensions},
Phys. Rev. Lett. {\bf 54}, No. 13 (1985) 1354-1356.}
\nref\DOU{see for instance M. Douglas, {\it The two-matrix model}, in 
{\it Random Surfaces and Quantum Gravity}, O. Alvarez,
E. Marinari and P. Windey eds., NATO ASI Series {B:} Physics Vol. {\bf 262} (1990).} 
\nref\CRI{J. Bouttier, P. Di Francesco and E. Guitter, {\it Critical and tricritical
hard objects on bicolourable random lattices: exact solutions}, J. Phys. A: Math. Gen.
{\bf 35} (2002) 3821-3854, arXiv:cond-mat/0201213.}

\vfill
\eject
\newsec{Introduction}

\subsec{Motivations}

In the recent years, a number of advances have related the combinatorics of
various classes of planar maps to that of decorated plane trees. More precisely, 
bijections based on appropriate cutting procedures were found between maps with 
prescribed vertex valences and so-called blossom trees [\xref\SCH-\xref\CENSUS], 
as well as between maps with prescribed face valences and so-called labeled mobiles
[\xref\CS,\xref\MOB], which are simply trees made of polygonal plaquettes. 
In this latter formulation, the labeling of the tree may be 
thought of as a one-dimensional embedding of its vertices. This allows [\xref\CS,\xref\GEOD]
for relating universal statistical properties of generic large random maps, known in the physics 
language as two-dimensional Quantum Gravity (2DQG)[\xref\QGRA,\xref\DGZ], to universal properties 
of large random embedded trees, captured in limiting models 
of embedded continuous random trees such as the Brownian Snake \BS\ or the Integrated 
SuperBrownian Excursion (ISE) in one dimension \ISE.
\fig{An example of a 2D Lorentzian map (a) with {\it vertices} of even valences and
a vertical shell structure, with slices delimited by dashed lines. For each slice, an oriented 
line enters from the left and exits to the right. The lines are non crossing but may
come into (possibly multiple) contacts,
giving rise to the (circled) vertices of the map. At each $2 p$-valent vertex, we erase (b) the $p-1$
(dashed) upper-rightmost outgoing edge halves, thus forming a tree (c) which can be planted
at its lower-rightmost edge.}{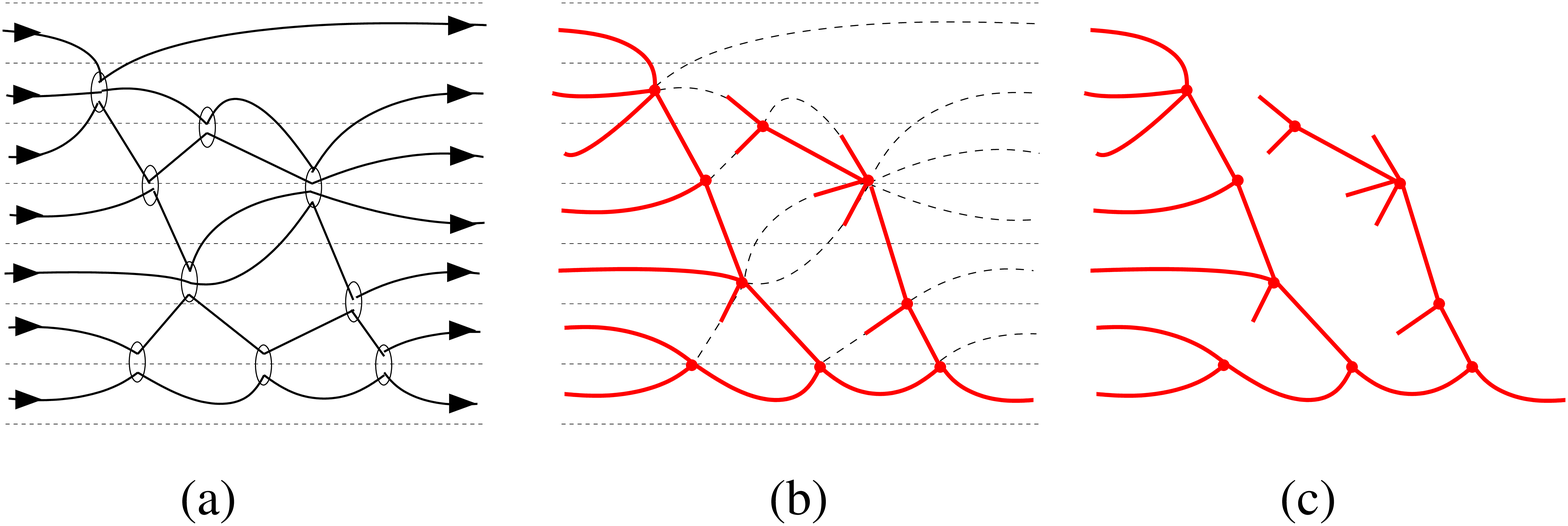}{13.5cm}
\figlabel\lorenty
\fig{An example (a) of a 2D Lorentzian map (dual to that of fig.1), made of polygonal {\it faces}
of even valences, and with a regular shell structure in the vertical direction. In each $2 p$-valent 
face, we pick the $p$ lower-leftmost vertices and join them into a (shaded) polygonal plaquette (which
reduces to a segment when $p=2$). The collection of these plaquettes forms a mobile (c), planted
at the bottom vertex.}{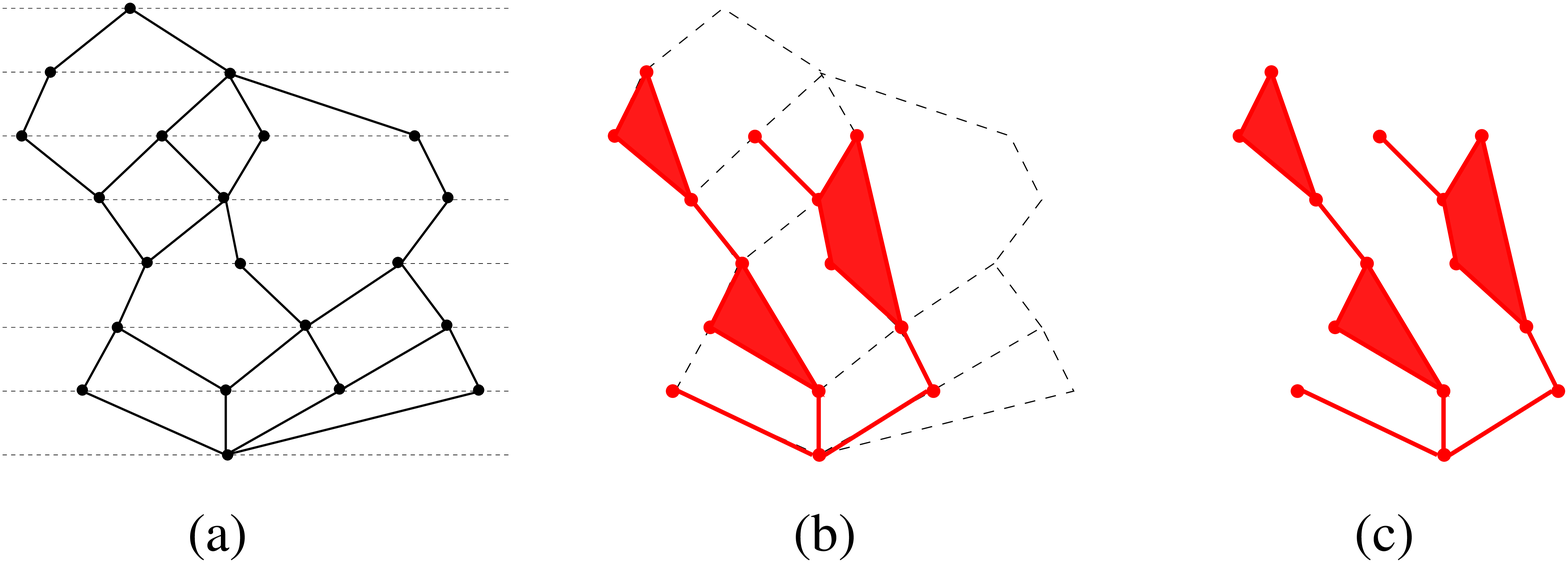}{13.5cm}
\figlabel\lorentx
A simpler version of this equivalence concerns maps describing the so-called two-dimensional 
Lorentzian Gravity (2DLG), i.e. random maps with a regular shell structure in one 
direction [\xref\AL,\xref\LORGRA].
As illustrated in figs.\lorenty\ and \lorentx, these are in one-to-one correspondence 
with plane trees without decorations (fig.\lorenty), 
or with unlabeled mobiles (fig.\lorentx). The universal large scale properties of 
generic Lorentzian random maps are now 
simply captured by that of a limiting model of continuous trees, the Brownian Continuous Random Tree 
(CRT) of Aldous \ALDOUS. 

The above equivalences may also be extended to maps with matter degrees of freedom, 
namely carrying configurations of statistical models such as the Ising model \BMS\ or hard
particles [\xref\CHP,\xref\HARGRA]. In the language of Quantum Gravity, these models may be fine-tuned so as to reach new critical points at which their
large scale statistical properties follow different laws than that of generic maps. 
These laws fall into universality classes characterized in particular by their order
of multicriticality.  When transposed into the tree language, this should 
give rise to natural multicritical generalizations of
the one-dimensional Brownian Snake or ISE, as well as of the CRT. 

A first family of multicritical points of Quantum Gravity may be
reached by considering maps with weights depending on the vertex valences and by fine
tuning these weights \ONEMATH. This allows to attain a $k$-th order multicritical point for
any integer $k\geq 2$, both
in the standard 2DQG framework and in the simpler 2DLG. Via the tree correspondence,
we expect similar $k$-th order multicritical points to arise in various fine-tuned
limits of large tree models. Some properties of a 
$k$-th order multicritical ISE might be inferred from ref.\GEOD\ by interpreting the multicritical 
two-point scaling function derived there as the average profile of a ``multicritical" 
continuous random tree embedded in the positive half-line.
Similarly, some properties of a $k$-th order multicritical CRT may be in principle read off the results 
of ref.\LORGRA\ for multicritical 2DLG.  A general understanding of multicritical ISE and CRT 
is however still lacking. 

The aim of this paper is to provide in the simpler case of unlabeled trees an explicit construction 
of a limiting model of Multicritical Continuous Random Tree, hereafter referred to as MCRT.
More precisely, we will derive basic properties of a $k$-th order multicritical continuous
random tree (denoted MCRT${}_k$) from suitable scaling limits of discrete models of
planted plane trees with properly fine-tuned vertex weights. These properties are expected 
to be universal and are used in a second step as guidelines for an independent self-consistent 
definition of the MCRT. 

Beyond the above-mentioned connection with 2DLG, another interest of the MCRT is that
it should also describe the universal properties of trees carrying particles subject to 
proper exclusion rules, such as that of ref.\HARGRA, and tuned to be at their multicritical
points (generalized Lee-Yang edge singularities). In a discrete formulation, these trees
may indeed be transformed into trees without particles under a redistribution of
the weights on the vertices. The critical points of such trees with particles will therefore
lie in the very same universality classes of MCRTs as those constructed here.

The paper is organized as follows. In sect.1.2 below, we present a short summary of
the main basic properties of the CRT, following Aldous [\xref\ISE,\xref\ALDOUS]. Section 2 is devoted
to a construction of the MCRT from discrete models of trees. These models are defined
in all generality in sect.2.1 and various statistical properties are introduced, such 
as the average profile and the so-called $m$-point correlation functions. 
We then discuss in sect.2.2 the emergence of critical and multicritical points in these models. 
The continuum scaling limits of large trees at these multicritical points
are investigated in sect.2.3 where we derive the average profile of the MCRT, 
as well as its ``history" distribution, describing the scaling limit of $m$-point correlations 
through the branching structure of the common genealogical tree of these points 
(hereafter called history tree). 
All these distributions are given explicitly via integral formulas which may also be interpreted in terms
of fractional derivative couplings at branching points corresponding to the most recent
common ancestors of the marked points. A number of properties of the MCRT are discussed in sect.2.4, 
such as the moments of its average profile and the relative weights of the possible branching 
structures for history trees.
Section 3 is devoted to the derivation {\it ab initio} of the very same properties of the MCRT 
without reference to a discrete model. The first step (sect.3.1) consists in using invariance principles
to obtain, with a minimal number of assumptions, a general form for the average profile
and history distributions of the MCRT. The precise value of these quantities is
derived by solving a set of consistency relations linking the correlations between $m$ marked
points to those with $m+1$. The net result is a fractional differential equation for
the average profile. This equation is transformed in sect.3.2 into an ordinary 
differential equation, which is solved in sect.3.3. The final result is shown 
in sect.3.4 to coincide with the explicit formula arising from the discrete approach of section 2.
A few concluding remarks are gathered in section 4.

\subsec{CRT: main properties}

Before we proceed with our construction, it is useful to first recall a few known facts
about Aldous' Brownian CRT [\xref\ISE,\xref\ALDOUS]. This CRT can be understood as the (weak) 
$N\to \infty$ limit of discrete planted
plane trees conditioned to have size $N$ (measured, say by the number of leaves),
with all edges scaled to have length $1/\sqrt{2N}$.
The properties of the CRT may be encoded into its average profile $\rho(x)$, which is the
limiting density of leaves at a given (rescaled) generation $x$ in the tree
\foot{We could as well consider the limiting density of edges at a given generation $x$
and we would find the same profile, up to a global normalization factor. We could also consider trees 
conditioned to have a fixed number $N'$ of {\it edges} rather than leaves, in which case the same 
limiting density is recovered provided edge lengths are now scaled by a factor  $1/\sqrt{N'}$}. 
The latter reads:
\eqn\rhoCRT{\rho(x)=x\, e^{-{x^2\over 2}}}
and is universal up to a possible rescaling $x\to \Lambda x$.
\fig{An example of continuous history tree of the CRT with $m=4$ distinguished leaves, characterized 
by its shape (planted binary plane tree with labeled leaves) and its $2 m-1=7$ branch lengths $x_j$. 
The associated probability density, as given by eq.(1.3),  is a function of the number $p=m-1=3$ of 
branching points and of the total length $x=\sum_j x_j$ only.}{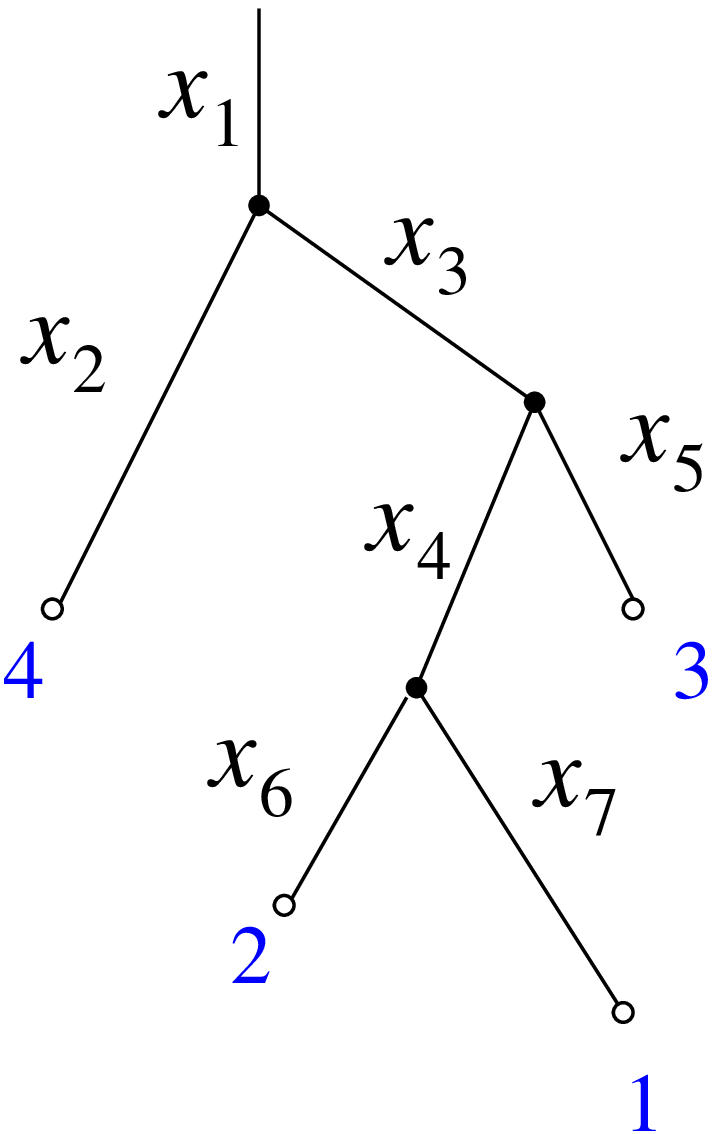}{4.cm}
\figlabel\binary
Correlations between points on the tree may be measured by marking, say $m$ distinguished leaves
chosen uniformly at random on the tree and considering the subtree spanned by the root and these 
leaves, i.e. their common genealogical tree formed by the union of their ancestral lines. 
This genealogical tree, or {\it history tree} $H$  is characterized by its {\it shape} $S$, 
which is a planted plane tree encoding the relative ordering of the most recent common ancestors
of the marked leaves, and by its (continuous) branch lengths recording the (rescaled)
number of generations between the most recent common ancestors or the marked leaves. 
Returning to discrete trees of size $N$, we note that the probability for three or more leaves to have
the same most recent common ancestor tends to zero as $N\to\infty$. As a consequence, 
in the continuous limiting CRT,
the only shapes with a non vanishing measure are {\it binary} plane trees whose leaves carry
the original marked leaves. An example of such an admissible history tree is depicted in fig.\binary.
For fixed $m$, there are $(2m-2)!/(m-1)!$ possible shapes (including the labeling 
of the distinct marked leaves). An important property of the CRT is that all these shapes turn out
to be equiprobable, with probability
\eqn\probasha{w(S)={(m-1)!\over (2 m-2)!}\ .}
More precisely, the history probability density for a given history tree $H$ depends only
on its number $p=m-1$ of branching points and on the {\it sum} of all its $2m-1$ (rescaled) 
branch lengths $x_j$, $j=1,\ldots,2m-1$ (see fig.\binary). It reads:
\eqn\histoCRT{\rho(H)= \left({1\over 2}\right)^p \rho\left(\sum_j x_j\right)\ ,}
with $\rho(x)$ as in eq.\rhoCRT. Note that the prefactor $(1/2)^p$ 
is an artefact of planarity and may be suppressed by considering non-planar trees, i.e.
by recording in a unique shape the $2^p$ shapes equivalent under arbitrary interchanges of the two 
descending subtrees around each vertex. Note finally that $w(S)$ is recovered by integrating
$\rho(H)$ over the $2m-1$ branch lengths $x_j$, which amounts to evaluating the $(2m-2)$-th moment of
$\rho(x)$.

\newsec{MCRT from discrete models}

\subsec{Discrete models}

The simplest approach to MCRT is via discrete models of random trees.
In this section, we shall concentrate on {\it planted plane trees}, i.e.
trees with a distinguished leaf (the {\it root}), and with vertices of arbitrary
but uniformly bounded valences. For $i\geq 1$, each $(i+1)$-valent vertex gives rise to $i$ 
{\it distinguished} descendent subtrees and receives a {\it weight} $g_i$ (with the
convention that $g_i=0$ for $(i+1)$ larger than the maximal allowed vertex valence).
We shall be interested in the ensemble ${\cal E}_N$ of 
trees with {\it fixed size}, defined as the total number $N$ of {\it leaves} in the tree. 
Each tree ${\cal T}$ therefore receives a total weight $\mu({\cal T})$ equal to
\eqn\weight{\mu({\cal T})={1\over Z_N(\{g_i\})} \prod_{i\geq 1\atop n_i({\cal T})>0} g_i^{n_i({\cal T})}\ ,}
where $n_i({\cal T})$ stands for the number of $(i+1)$-valent vertices in ${\cal T}$ ($i\geq 1$), 
and where the normalization factor, or {\it partition
function} $Z_N(\{g_i\})$ is chosen so that
\eqn\normZ{\sum_{{\cal T}\in {\cal E}_N}\mu({\cal T})=1\ .}
The MCRT will be associated to various large $N$ scaling limits of this ensemble 
${\cal E}_N$ for suitably fine-tuned weights.
\fig{The recursive formula for the tree generating function $T$ in the grand-canonical ensemble 
${\cal E}$.  The trees are decomposed according to the environment of the vertex attached to the root.
If this vertex is a leaf, it receives a weight $\lambda$. If the valence of this vertex is $i+1$
($i\geq 1$), we get a weight $g_i$ for the vertex and a factor $T$ for each of the $i$ 
descendent subtrees. }{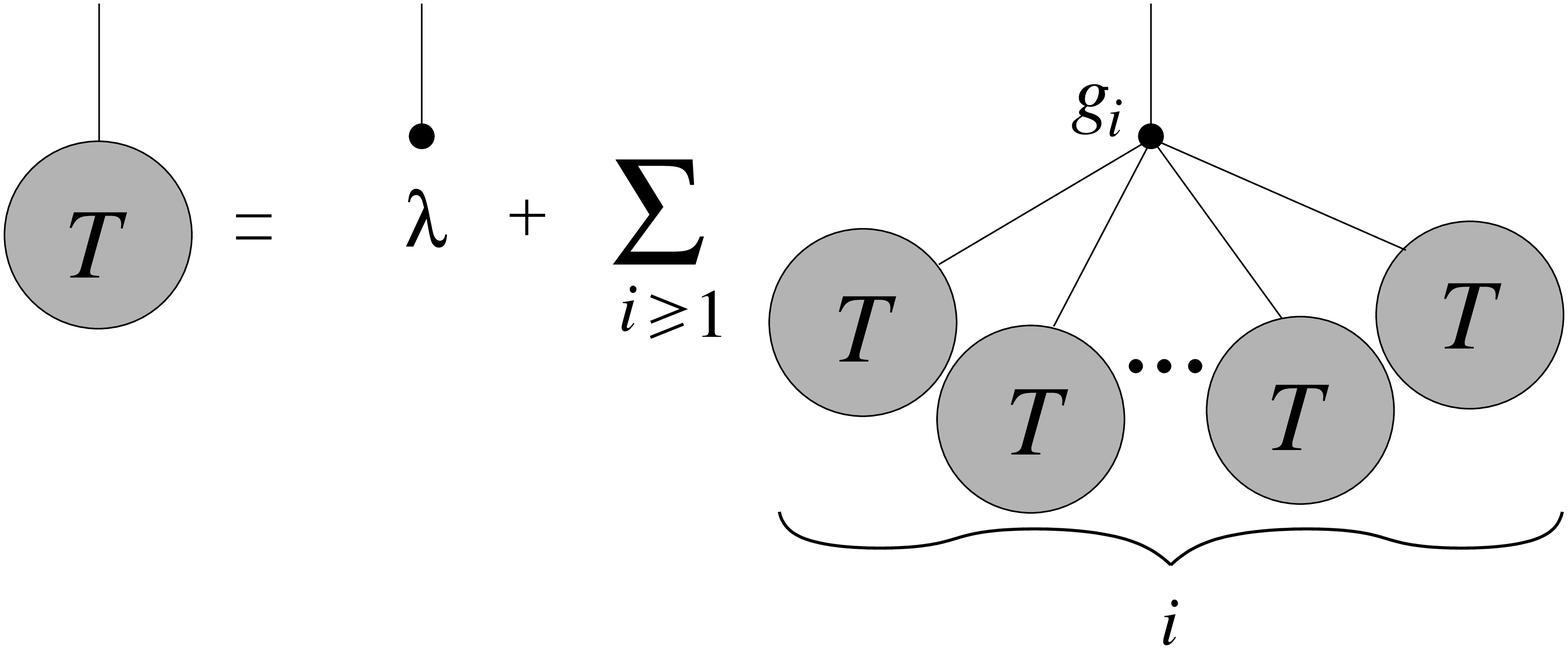}{8.cm}
\figlabel\recur
For convenience, it is also useful to consider the related, {\it grand canonical}, 
ensemble ${\cal E}$ of trees of arbitrary size,
but now with an extra weight $\lambda$ per leaf. In this new ensemble, each 
tree ${\cal T}$ receives an un-normalized weight 
\eqn\weightgc{\nu({\cal T})=\lambda^{N({\cal T})}\, \prod_{i\geq 1\atop n_i({\cal T})>0} 
g_i^{n_i({\cal T})}\ ,}
with $N({\cal T})$ its total number of leaves.
Introducing the corresponding generating function 
\eqn\gfT{T\equiv \sum_{{\cal T}\in {\cal E}}\nu({\cal T})=\sum_{N\geq 1} \lambda^N Z_N(\{g_i\})\ ,}
we have the recursive formula illustrated in fig.\recur:
\eqn\recurT{T= \lambda + f(T)\ , \qquad f(T)\equiv\sum_{i\geq 1}g_i T^i\ ,}
obtained by inspecting the environment of the vertex attached to the root. 
Note that, as valences are bounded, $f$ is polynomial in $T$.
Equation \recurT\ determines uniquely $T(\lambda)$ as a formal power series
of $\lambda$ whose coefficients are polynomials of the $g_i's$, and which is the functional
inverse of $\lambda(T)\equiv T-f(T)$ defined in a vicinity of $T=0$. 

The knowledge of $T(\lambda)$ gives access to the partition function $Z_N$ via the
contour integral (encircling the origin)
\eqn\Ttozn{Z_N=\oint {d\lambda \over 2 i \pi \lambda^{N+1}}T(\lambda)\ .}
More generally, we may consider the {\it average profile} $\rho_N(L)$ of trees in ${\cal E}_N$, 
equal to the weighted sum over trees of size $N$ with a marked leaf at distance $L\geq 1$ 
from the root. The average profile reads 
\eqn\rhonl{\rho_N(L)={1\over N Z_N}\oint {d\lambda \over 2 i \pi \lambda^{N+1}}
\left\{f'(T(\lambda))\right\}^{L-1} \lambda\ .}
Indeed, the marked leaf defines a unique minimal path from the root passing through  
$L-1$ inner vertices. At each of these $L-1$ vertices, the choice of the outgoing edge 
produces the weight $f'(T(\lambda))$. Finally the marked leaf receives a weight $\lambda$. 
The prefactor $1/(N Z_N)$ ensures the correct normalization $\sum_{L\geq 1} \rho_N(L)=1$.
\fig{Starting from a marked tree (a), here with $m=4$ marked leaves (filled circles), 
we form (b) the union of all minimal paths (thick edges) from the root to these leaves.
Dropping all other edges and vertices, this defines a marked history tree (c) whose branching
points (encircled here) correspond to pairwise most recent common ancestors. 
To each branch is associated a length $L_j\geq 1$. Here $L_5=1$, 
$L_1=2$, $L_2=3$, $L_4=L_6=4$ and $L_3=6$.}{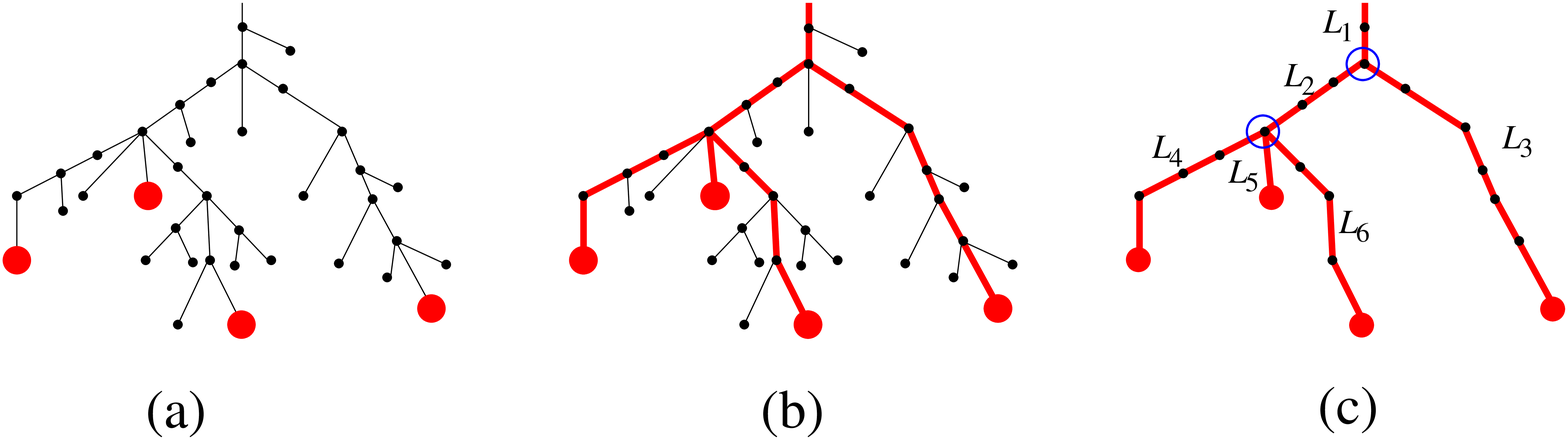}{13.5cm}
\figlabel\histtree
We may also consider $m$-point correlation functions, i.e. averages over trees with a 
number $m\geq 2$ of marked leaves chosen uniformly and independently at random among all leaves
different from the root. For each such marked tree ${\cal T}$, we
may define its {\it history tree} ${\cal H}$ as the union of all minimal paths from the root to 
the $m$ marked leaves (see fig.\histtree).
This union of {\it ancestral lines} indeed forms a tree whose branching points correspond to most recent
common ancestors of the marked leaves. 
For any history tree ${\cal H}$, let us denote by $p_i$, $i\geq 2$ its number of $(i+1)$-valent (on 
${\cal H}$) branching points and $p_0$ its number of leaves (other than the root). 
The branching points of ${\cal H}$ divide it into a number $n$ of ``branches", with 
\eqn\valn{n=p_0+\sum_{i\geq 2} p_i=1+\sum_{i\geq 2} i p_i\ .} 
These branches have lengths $L_j\geq 1$, $j=1,\ldots,n$. Note that the history tree must be supplemented
with the data of the positions of the $m$ marked leaves: this will be called a {\it marked history
tree} (still denoted ${\cal H}$ by a slight abuse of notation). By construction, all the $p_0$ 
leaves must be marked at least once. This implies the inequality 
\eqn\ineq{m\geq p_0=1+\sum_{i\geq 2} (i-1)p_i\ .}
A marked history tree will be called {\it non-degenerate} if no two marked leaves 
are equal. This corresponds to having $m=p_0$.  It is easily seen that 
non-degenerate history trees dominate in the large $N$ limit. 
The weighted sum over all marked trees sharing the same marked history ${\cal H}$ reads
\eqn\rhommm{\rho_N({\cal H})={1\over N^m Z_N}\oint {d\lambda \over 2 i \pi \lambda^{N+1}}
\left\{f'(T)\right\}^{\sum\limits_j (L_j-1)} \prod_{i\geq 2} \left\{{f^{(i)}(T)
\over i!}\right\}^{p_i} \lambda^{p_0}}
by a straightforward generalization of the above argument leading to eq.\rhonl. 
The factors $f^{(i)}(T)/i!$ account for the choices of the $i$ descending subtrees 
at $(i+1)$-valent branching points. 
The prefactor $1/(N^m Z_N)$ ensures the correct normalization $\sum_{\cal H} \rho_N({\cal H})=1$,
where the sum runs over all (possibly degenerate) histories with $m$ marked leaves.

\subsec{Critical and multicritical points}

For fixed $g_i$'s, the large $N$ properties of the ensemble ${\cal E}_N $ are related to 
the properties  
of the ensemble ${\cal E}$ when $\lambda\to \lambda_c$, where $\lambda_c$ is the
radius of convergence of $T(\lambda)$. It is obtained by first writing $\lambda'(T)=0$, which determines
$T=T_c$, and then identifying $\lambda_c=\lambda(T_c)$. This equivalently amounts to the system
of equations\foot{Note that, in general, this system has more than one solution and
we must select among those the actual radius of convergence $\lambda_c$.}
\eqn\radcon{T_c=\lambda_c\,+\,f(T_c)\qquad {\rm and} \qquad 1= \, f'(T_c)\ .}
For a generic choice of the $g_i$'s, we have $f''(T_c)\neq 0$, which implies, by a Taylor expansion
of $\lambda$ around $T_c$, that $(\lambda_c-\lambda)\propto (T_c-T)^2$, corresponding to a generic
square root singularity of $T$ as a function of $\lambda$.
A higher multicritical behavior of order $k>2$ may be reached by tuning the $g_i$'s
so as to enforce the relations
\eqn\finetun{f''(T_c)=\cdots=f^{(k-1)}(T_c)\ , \qquad f^{(k)}(T_c)\neq 0\ .}
A Taylor expansion of $\lambda$ around $T_c$  now leads to the behavior
\eqn\mulbe{{\lambda_c-\lambda\over \lambda_c}\sim A \left({T_c-T\over T_c}\right)^k
\ , \qquad A\equiv (-1)^k {f^{(k)}(T_c)\over k!}{T_c^k\over \lambda_c}\ ,}
leading to a $k$-th root singularity of $T$ as a function of $\lambda$.
A minimal realization of conditions \finetun\ above consists in taking
\eqn\minf{f(T)= {k T -1 + (1-T)^k\over k}}
in which case, $T_c=1$, $\lambda_c=1/k$ and $A=1$.  This corresponds to the
choice $g_i=g_i^*$
\eqn\ming{g_1^*=0 \quad {\rm and} \quad g_i^*={(-1)^i \over k}{k\choose i}\ {\rm for}\ i=2, \ldots, k}
while $g_i^*=0$ for $i>k$. 

In the following, we shall consider the general case of a function $f(T)$ satisfying the
$k$-th order multicriticality conditions \finetun. However, as the values of $T_c$, $\lambda_c$ and $A$ 
are non-universal, we decide to fix them, without loss of generality, to the same values as in the 
minimal case \minf\ above, namely  $T_c=A=1$, $\lambda_c=1/k$.
This amounts to taking a function $f(T)$ satisfying
\eqn\extminf{f(T)= {k T -1 + (1-T)^k\over k}+{\cal O}\left((1-T)^{k+1}\right)}
and making sure that the radius of convergence of $T(\lambda)$ remains at 
$\lambda=\lambda_c=1/k$. 
Writing $f(T)=\sum_i g_i T^i$, the property \extminf\ is obtained by imposing 
$k+1$ relations on the $g_i$'s. In particular, reaching a $k$-th order multicritical point
requires to consider planted plane trees with at least $k-1$ different values for the inner 
vertex valences strictly larger that $2$. For $k>2$, this moreover forces some of the $g_i$'s to 
be {\it negative}. The minimal choice \minf\ amounts to picking the lowest possible
vertex valences, namely $3,\ldots,k+1$.

\subsec{Continuum limit}

In the light of the discussion of previous section, we may now consider the large $N$ behavior
of the functions $Z_N$, $\rho_N(L)$ and $\rho_N({\cal H})$ of eqs.\Ttozn, \rhonl\ and \rhommm\
at a $k$-th order multicritical point, i.e. for fixed values of the $g_i$'s that ensure
eq.\finetun. We will be led to various rescalings
of the trees, eventually allowing for a continuous formulation describing a
{\it $k$-th order multicritical continuous random tree} of fixed size $1$, hereafter 
referred to as the MCRT${}_k$. 
\fig{The contour of integration in eq.(2.17) for the $T$ variable in the complex plane. We deform
the contour on the left so as to approach the large $N$ saddle point value $T_c$. The approach to $T_c$ is
tangent to the lines with angles $\pm \pi/k$ corresponding the real values of $\lambda(T)$
with a minimum at $T_c$. At large $N$, the integral is dominated by the vicinity of
$T_c$ (zoomed here) and amounts to an integral over a real $\xi$ variable as shown,
with $\omega=e^{i\pi{k-1\over k}}$.}{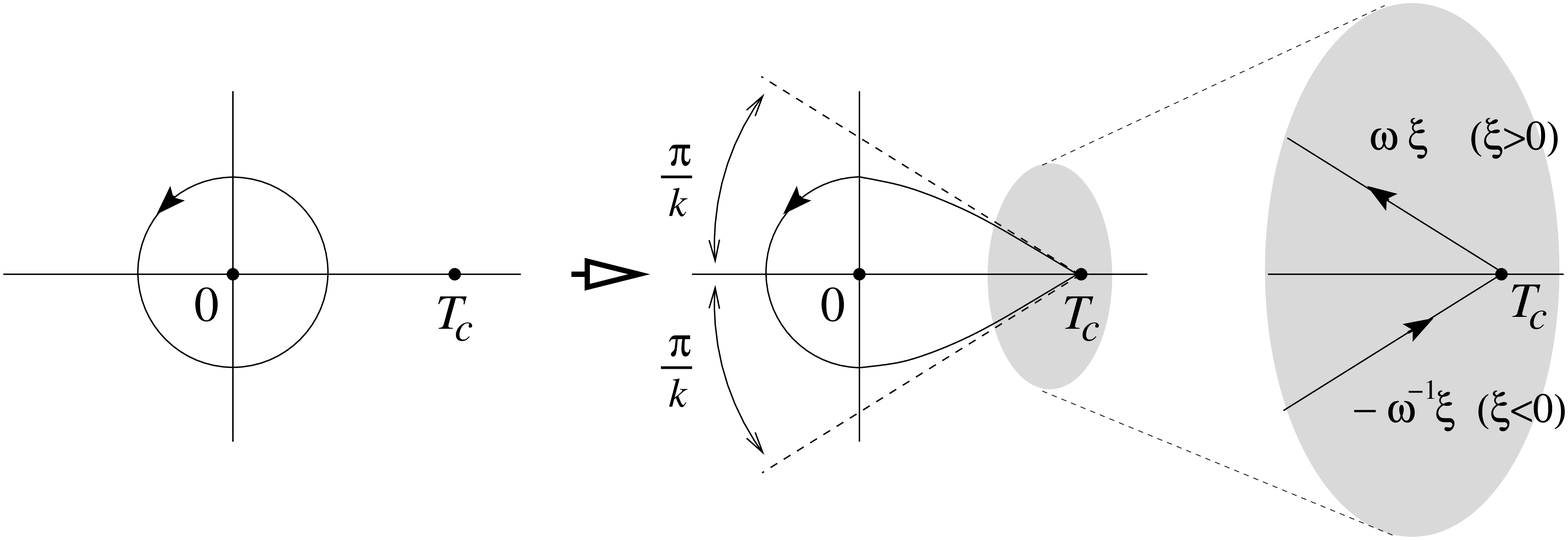}{13.cm}
\figlabel\contour
The integral \Ttozn\ for $Z_N$ may be evaluated by first changing variables from $\lambda$
to $T$, namely
\eqn\ltoT{Z_N=\oint {dT\over 2 i \pi}\, T \, {\lambda'(T)\over \lambda(T)} 
e^{-N {\rm Log}\lambda(T)}
\ .} 
Assuming the generic form \extminf\ for $f(T)$ and writing $\lambda(T)=T-f(T)$, 
this integral is dominated at large $N$ 
by the saddle point $T=T_c=1$ around which eq.\mulbe\ holds
with $\lambda_c=\lambda(T_c)=1/k$. This suggest to deform the contour of integration for $T$ 
as displayed in fig.\contour. In the vicinity of $T_c$, this amounts to switching to a real variable 
$\xi$ such that 
\eqn\xitoT{T=T_c\left(1+\omega^{{\rm sgn}(\xi)} {\vert \xi \vert \over (kN)^{1\over k}}\right)}
where $\omega=e^{i\pi {k-1\over k}}$, leading to 
\eqn\defxi{\lambda\sim\lambda_c\left(1+ A {\vert \xi\vert^k\over kN}\right)}
with $A=1$.
The precise choice of phase $\omega^{{\rm sgn}(\xi)}$ ensures that $\lambda$ be real and minimal
at $T_c$ on the new contour, while the factors of $N$ have been tuned so that the argument of the 
exponential term in \ltoT\ be of order zero. The large $N$ equivalent of $Z_N$ 
is obtained by integrating $\xi$ on the real
axis, leading to
\eqn\explitzn{\eqalign{Z_N &\sim {k^N \over 2 i \pi N} \int_{-\infty}^{\infty}
d\xi \ {\rm sgn}(\xi) \vert\xi\vert^{k-1} e^{-{\vert \xi \vert^k\over k}}\left(
1+\omega^{{\rm sgn}(\xi)}{\vert \xi \vert \over (kN)^{1\over k}}\right)
\cr &= {k^{N-{1\over k}} \over \pi N^{1+{1\over k}} } 
{\rm Im}(\omega)
\int_0^{\infty}
d\xi \xi^k e^{-{\xi^k\over k}}
\cr &= {k^N \over \pi N^{1+{1\over k}}} \sin\left({\pi \over k}\right)
\Gamma\left(1+{1\over k}\right)\ .\cr}}
Repeating the above exercise for $\rho_N(L)$ as in eq.\rhonl, we simply have to take into
account the extra factor $(f'(T))^{L-1}$ with
\eqn\lfp{f'(T)\sim 1-(1-T)^{k-1}= 1+ \tau^{{\rm sgn}(\xi)} {\vert \xi \vert^{k-1}\over 
(kN)^{{k-1\over k}}}}
where $\tau=-(-\omega)^{k-1}=e^{i{\pi \over k}}$. In order to get a sensible scaling limit, we must take 
\eqn\lscal{L\sim x (kN)^{{k-1\over k}}\ .}
In our construction, the MCRT${}_k$ therefore corresponds to a 
limit of $k$-th order multicritical discrete trees 
in which edges are scaled to have length $(kN)^{-{k-1\over k}}$ rather than $1$. We shall refer
to $d_k=k/(k-1)$ as the {\it fractal dimension} of the MCRT${}_k$ ($N\sim L^{d_k}$). 
With the above redefinitions, the rescaled average profile
\eqn\defrho{\rho(x)\equiv\lim_{N\to \infty} (kN)^{{k-1\over k}} \rho_N\left(x (kN)^{{k-1\over k}}\right)}
takes the form
\eqn\intformu{\rho(x)= {1\over k^{1\over k} \Gamma\left({1\over k}+1\right)
\sin\left({\pi\over k}\right)} {\rm Im} \left\{ 
\int_0^\infty d\xi \xi^{k-1} e^{-{\xi^k\over k}+\tau \xi^{k-1} x} \right\}
\quad {\rm with} \quad \tau=e^{i{\pi \over k}}\ .}
\fig{Plot of the universal average profile $\rho(x)$ of eq.\intformu\ for 
$k=2,3,4$.}{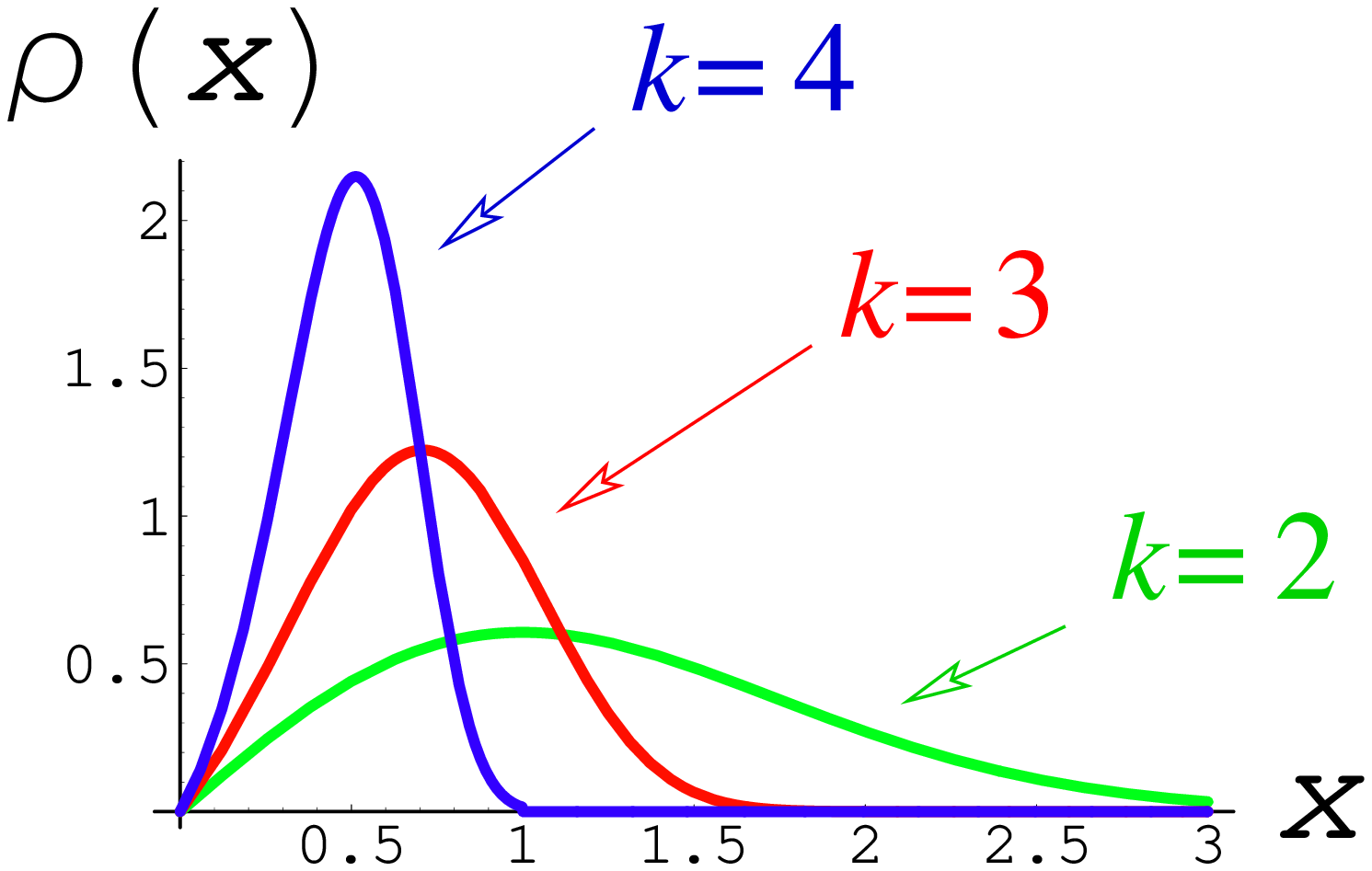}{7.cm}
\figlabel\rhoplot
\noindent This function is the {\it universal average profile} for all ensembles of $k$-th order
multicritical trees. It is represented in fig.\rhoplot\ for $k=2,3,4$. For $k=2$, eq.\intformu\ 
reduces to eq.\rhoCRT\ as expected, since the CRT is nothing but the MCRT${}_2$.

We may alternatively obtain an integral transform of $\rho(x)$ by considering the critical continuous
counterpart of the equation
\eqn\invrhonl{\sum_{N\geq 0} \lambda^N N Z_N \rho_N(L) = \big(f'(T(\lambda))\big)^{L-1} \lambda}
which is the grand-canonical equivalent of eq.\rhonl. Approaching the critical point via
\eqn\approcrit{\lambda={1\over k}e^{-\epsilon}\ , \quad  L={x\over \epsilon^{{k-1\over k}}}}
with $\epsilon \to 0$, we use $f'(T)\sim 1-\epsilon^{k-1\over k}$ and $T\to 1$ to rewrite 
the r.h.s. of eq.\invrhonl\ as $e^{-x}$. The sum in the l.h.s. tends to an integral over
a variable $M= \epsilon\, kN$ so that eq.\invrhonl\ turns into
\eqn\invrho{
{k^{{1\over k}} \sin\left({\pi \over k}\right)
\Gamma\left(1+{1\over k}\right) \over \pi}
\int_0^\infty {dM\over M^{1+{1\over k}}} e^{-{M\over k}} \left\{{M\over M^{k-1\over k}}
\ \rho\left({x\over M^{{k-1\over k}}}\right)\right\}=
e^{-x}}
where we have used eqs.\explitzn\ and \defrho. The variable $M$ may be viewed as the 
(continuous) size of the trees in the MCRT and the quantity 
\eqn\rescarho{\rho_M(x)\equiv {M \over M^{{k-1\over k}}}\rho\left({x\over M^{{k-1\over k}}}\right)}
is nothing but the average profile for the MCRT${}_k$ with fixed size $M$,
normalized by $\int_0^\infty \rho_M(x)dx=M$.

We may finally derive the scaling limit of the $m$-point correlation function corresponding to
a given marked history tree.
Rewriting eq.\rhommm\ as 
\eqn\rhommmbis{\rho_N({\cal H})={1\over N^m Z_N}\oint {d\lambda \over 2 i \pi \lambda^{N+1}}
\big(f'(T)\big)^{\sum\limits_j (L_j-1)} \prod_{i\geq 2} \left\{{\lambda^{i-1}f^{(i)}(T)
\over i!}\right\}^{p_i} \,\lambda}
and comparing with $\rho_N(L)$, we simply have to substitute $(L-1) \to \sum_j (L_j-1)$ and to
insert extra factors of the form $\lambda^{i-1}f^{(i)}(T)/i!$. These factors scale as
\eqn\extrafac{\left\{{\lambda^{i-1}f^{(i)}(T)
\over i!}\right\}\sim {(-1)^i \over k^i}{k \choose i} (1-T)^{k-i}={1\over k^{i{k-1\over k}}} 
{(-1)^i \over k}{k \choose i} 
\left(-{\omega^{{\rm sgn}(\xi)}\vert \xi \vert\over N^{{1\over k}}}\right)^{k-i}}
for $i=2,\ldots,k$, while for $i>k$, they tend to constants depending on the $g_i$'s only.
Again, taking the scaling $L_j\sim x_j (kN)^{{k-1\over k}}$ for $j=1,\ldots n$ (with $n$ as in eq.\valn), 
let us define the rescaled correlation function
\eqn\defrhoH{\rho(H)\equiv\lim_{N\to \infty} (kN)^{n{k-1\over k}} \rho_N({\cal H})\ .}
Here the notation $H$ stands for the continuous history tree corresponding to ${\cal H}$
defined as follows.
The tree $H$ has the same branching structure as ${\cal H}$, but the $n$ branches are
replaced by segments of real lengths $x_j>0$. 
Collecting the factors of $N$ in eq.\rhommmbis, we get an overall
factor $N^{-\alpha}$ in \defrhoH, with
\eqn\glog{\alpha=m-{1\over k}-n{k-1\over k}+\sum_{i=2}^k{k-i\over k}p_i
=(m-p_0)+{1\over k}\sum_{i>k}(i-k)p_i\ .}
\fig{A continuous history tree $H$ is characterized by its shape, namely a planted plane tree
with branching points of valence less than $k+1$ (here $k\geq 4$) and by the real lengths $x_j>0$ 
of its branches. The leaves are distinguished and labeled from 
$1$ to $m$ (here $m=7$).}{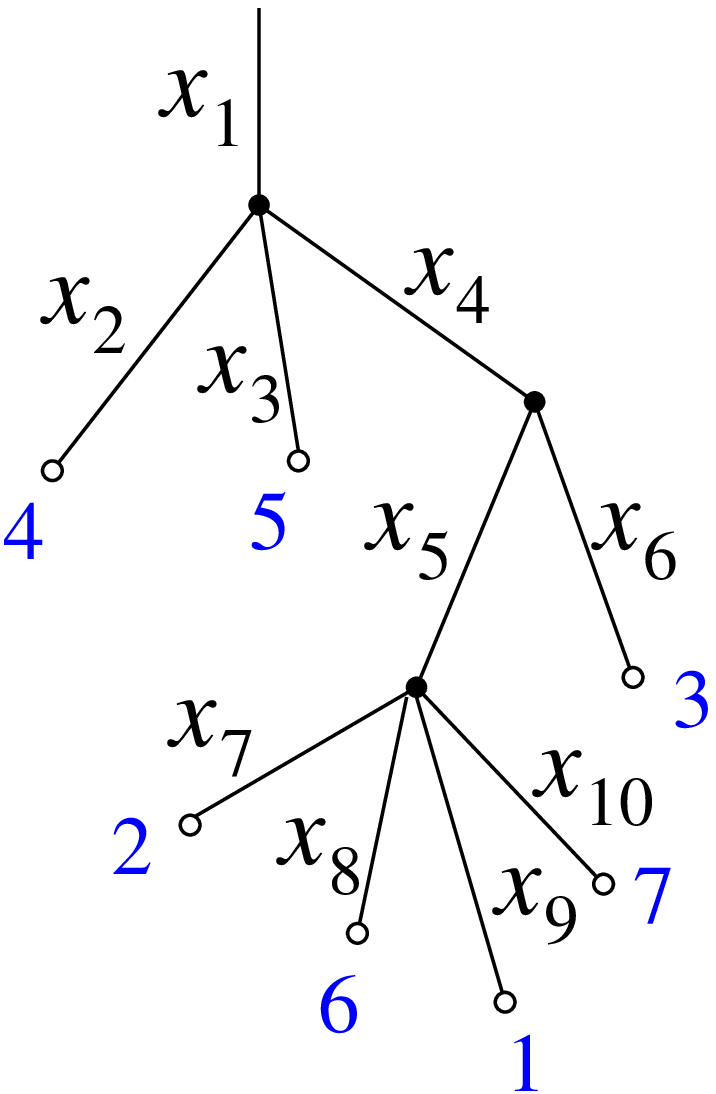}{4.cm}
\figlabel\history
\noindent The exponent $\alpha$ is always non-negative from eq.\ineq, and vanishes only
if $p_i=0$ for all $i>k$ and $m=p_0$. This selects marked history trees which are 
non-degenerate and whose branching points have valences less or equal to $k+1$.
In other words, the discrete histories surviving in the continuum limit at a $k$-th order multicritical
point are those with branching points with valences $3,4,\ldots (k+1)$ only, and
all leaves but the root marked exactly once. From now on, we shall always consider 
such histories. As displayed in fig.\history, the 
associated continuous history $H$ is entirely characterized by the data (i) of a {\it shape},
namely a planted plane tree with branching points of valence $3,4,\ldots (k+1)$ and with
{\it distinguished} leaves labeled $1,\ldots,m$, and (ii) of 
a collection of real lengths $x_j$ for its edges. We still denote by $p_i$, $i=2,\ldots k$ the number of its
$(i+1)$-valent branching points. This generalizes to arbitrary values $k\geq 2$ the binary structure of 
the CRT discussed in sect.1.2, recovered here for $k=2$.
The correlation $\rho(H)$ reads
\eqn\rhohcont{\rho(H)=
{\prod\limits_{i=2}^k (\mu_i)^{p_i} 
\over k^{1\over k} \Gamma\left({1\over k}+1\right)
\sin\left({\pi\over k}\right)} 
\ {\rm Im} \left\{
\int_0^\infty d\xi \xi^{k-1}\, \left({\xi\over \tau}\right)^{\sum\limits_{i=2}^k
(k-i)p_i} e^{-{\xi^k\over k}+\tau \xi^{k-1} \sum\limits_j x_j} \right\}}
where $\tau=e^{i{\pi\over k}}$ and 
\eqn\valmui{\mu_i\equiv{(-1)^i \over k} {k \choose i}\ .}
This gives the {\it universal history distribution} for the MCRT${}_k$, generalizing the
CRT history probability density \histoCRT. 
The formula \rhohcont\ may be interpreted as follows. First note that $\rho(H)$ only depends
on the total length $x\equiv \sum_j x_j$ of $H$. Furthermore, to each $(i+1)$-valent branching point of
$H$ is associated a weight $\mu_i$, whose value is universal and given by \valmui.
Note that these universal weights carry alternating signs for $k\geq 3$. Finally, in addition to these
numerical weights, each $(i+1)$-valent branching point contributes a factor
$(\xi/\tau)^{k-i}$ to the integrand. Note that for $i=k$, this contribution reduces to $1$. In particular,
in the case of a history $H$ with only $(k+1)$-valent branching points, the history distribution
$\rho(H)$ is simply proportional to $\rho(x)$. The expression \rhohcont\ may be recast in 
terms of {\it fractional derivatives} of $x$ as 
\eqn\rhohfrac{\rho(H)=\prod_{i=2}^k \left(\mu_i (-d)^{{k-i\over k-1}} \right)^{p_i} \rho(x)}
with $x\equiv \sum_j x_j$ and where $\rho(x)$ is the average profile given by eq.\intformu.
Here the fractional derivative $(-d)^{r\over k-1}$ for non-negative integer 
$r$ acts on a function $\varphi(x)$ with integral representation
\eqn\intphi{\varphi(x) = {\rm Im}\left\{\int_0^\infty d\xi\, {\tilde\varphi}(\xi) 
e^{-{\xi^k\over k}+\tau x \xi^{k-1}} \right\}}
as
\eqn\resintphi{(-d)^{{r\over k-1}}\varphi(x) = {\rm Im}\left\{\int_0^\infty d\xi\, 
\left({\xi\over \tau}\right)^r\,  
{\tilde\varphi}(\xi) e^{-{\xi^k\over k}+\tau x \xi^{k-1}} \right\}\ .}
This fractional derivative coincides with (minus) the usual derivative with respect to
$x$ when $r=(k-1)$, and satisfies the additivity property
\eqn\daddit{(-d)^{{r\over k-1}}(-d)^{{s\over k-1}}= (-d)^{{r+s\over k-1}}}
and the generalized Leibniz formula 
\eqn\Leibfor{(-d)^{r\over k-1} x \varphi(x)= x (-d)^{r\over k-1}\varphi(x) 
-{r\over k-1}(-d)^{{r\over k-1}-1} \varphi(x)\ .} 
The definition also extends to negative values of $r$ provided ${\tilde\varphi}(\xi)={\cal O}
(\xi^{-r})$ for $\xi\to 0$. In particular, for $r=-(k-1)$ and ${\tilde\varphi}(\xi)={\cal O}
(\xi^{k-1})$, we have 
\eqn\dtoint{(-d)^{-1}\varphi(x)=\int_0^\infty dy\ \varphi(x+y) .} 
Alternatively, we may avoid using fractional derivatives at the expense of introducing
$k-1$ basic (normalized) distributions $\sigma_j$, $j=2, \ldots, k$, defined as
\eqn\defsig{
\sigma_j(x)\equiv {k^{{j-1\over k}}\over \Gamma\left({k-j+1\over k}\right)
\sin\left((k-j+1){\pi\over k}\right)} {\rm Im} \left\{ 
\int_0^\infty d\xi \left({\xi \over \tau}\right)^{k-j} \xi^{k-1} 
e^{-{\xi^k\over k}+\tau \xi^{k-1} x} \right\}\ ,}
such that $\rho(x)=\sigma_k(x)$ and $\rho(H)\propto \mu_j\sigma_j(\sum_{i=1}^n x_i)$ if $H$ 
has a unique $(j+1)$-valent branching point.
The distribution associated with any history is then proportional to $d^m \sigma_j(x)/dx^m$ 
where $m$ and $k-j$ are the quotient and the remainder of the Euclidean division: 
$\sum_i (k-i)p_i=m(k-1)+k-j$.

To conclude this section, let us mention the generalization of eq.\invrho\ to arbitrary
histories:
\eqn\invhist{
{k^{{1\over k}} \sin\left({\pi \over k}\right)
\Gamma\left(1+{1\over k}\right) \over \pi}
\int_0^\infty {dM\over M^{1+{1\over k}}} 
e^{-{M\over k}} 
\left\{{M^{m}\over M^{n{k-1\over k}}} 
\ \rho\left({H\over M^{{k-1\over k}}}\right)\right\}= e^{-x} \prod_{i=2}^k (\mu_i)^{p_i} }
where $x=\sum_i x_i$ and where the notation $H/\Lambda$ means that all the $x_i$'s
of $H$ are replaced by $x_i/\Lambda$.
\fig{A plot of the $k=3$ fundamental distributions: (a) the average profile $\rho(x)$ and (b)
the distribution $\sigma(x)\equiv \mu_2 (-d)^{1\over 2}\rho(x)$. These distributions (solid lines) 
are compared with the corresponding discrete data (dotted lines) obtained from the minimal
model \minf\ for a number $N=200$ of leaves.}{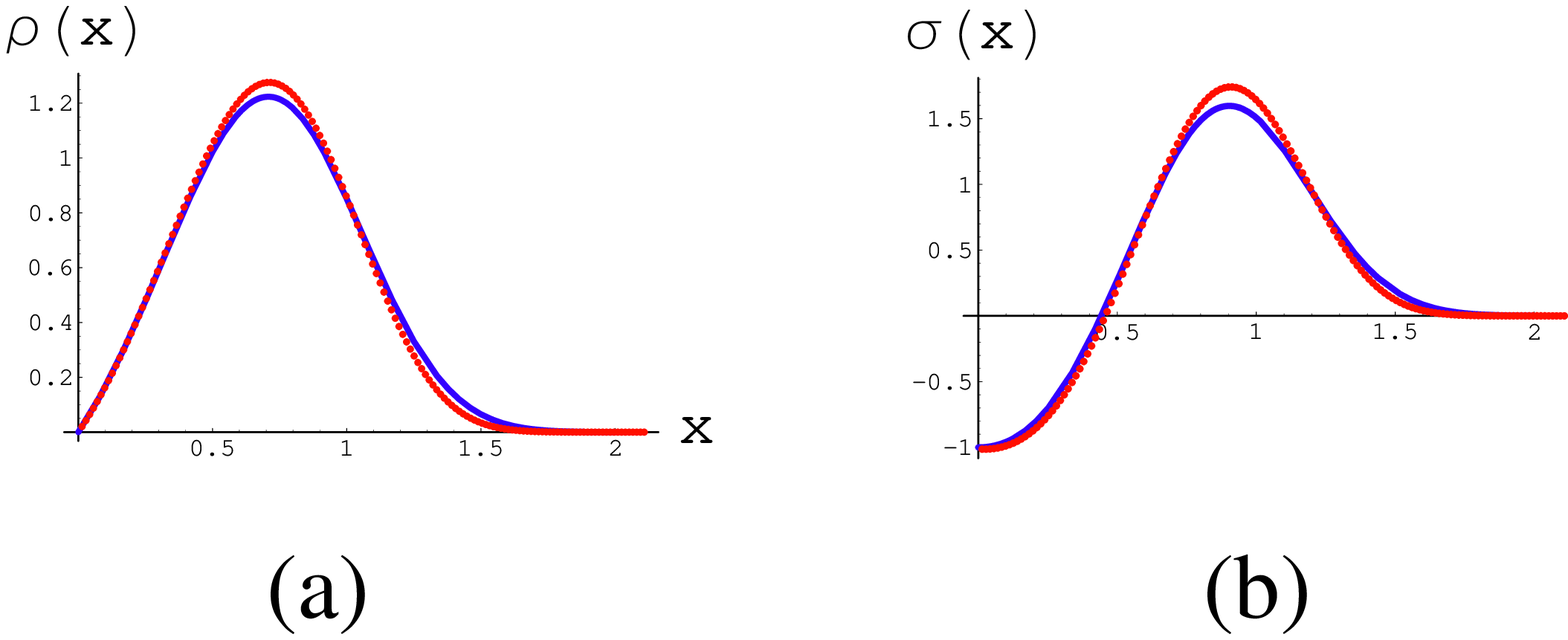}{13.cm}
\figlabel\discret
Finally, let illustrate the case $k=3$ by considering the two fundamental distributions:
the average profile $\rho(x)$ and the distribution $\sigma(x)\equiv \mu_2 (-d)^{1\over 2}\rho(x)$ 
corresponding to $m=2$ marked leaves and a total length $x=x_1+x_2+x_3$. These functions
are plotted in fig.\discret\ against their properly rescaled discrete counterparts (see eqs.
\defrho\ and \defrhoH) in the particular case of the minimal model, namely with tri- and
tetra-valent vertices weighted respectively by $1$ and $-1/3$, according to eq.\ming.
Note that $\sigma(x)$ is negative for small $x$, which shows that the MCRT corresponds to 
a signed measure. This is a generic property for all $k>2$. 

\subsec{A few properties of the MCRT${}_k$}

Having the exact form of the average profile and history distributions \intformu\ and \rhohcont, we
may infer a number of properties of the MCRT${}_k$.
A first question of interest is the large $x$ asymptotics of $\rho(x)$. 
A saddle point approximation shows that
\eqn\asymprho{\rho(x)\sim \gamma(k)\  x^{k\over 2}  e^{-{(k-1)^{k-1}\over k} x^k}\ .}
for some constant $\gamma(k)$. We recognize Fisher's law relating the exponent $\delta$ of
the large $x$ exponential decay $\rho(x)\sim e^{-a x^\delta}$ to the fractal dimension via
$\delta=d_k/(d_k-1)=k$. In particular, this behavior guarantees the existence of all moments of
$\rho(x)$:
\eqn\defmomen{\langle x^b \rangle_\rho \equiv \int_0^\infty dx\ x^b\ \rho(x)\ .}
These may be readily computed from eq.\invrho\ upon multiplication by $x^b$ and integrating
over $x$.
We find immediately
\eqn\rhomom{
{k^{{1\over k}} \sin\left({\pi \over k}\right)
\Gamma\left(1+{1\over k}\right) \over \pi}
\int_0^\infty {dM\over M} M^{{k-1\over k}(b+1)} e^{-{M\over k}} \ \langle x^b \rangle_\rho =
b! }
henceforth
\eqn\momrho{\langle x^b \rangle_\rho
={b! \over  k^{b{k-1\over k}}}{\Gamma\left({k-1\over k}\right)\over \Gamma\left((b+1){k-1\over k}\right)}\ .}

Beyond the properties of the average profile, we may compare the relative contributions
of various histories as follows. Recall that a continuous history $H$ is given by two types of data: 
(i) its shape $S$ recording
its branching structure, characterized by the numbers $p_i$ and (ii) the lengths $x_j$ of
its branches.  For a given number $m$ of marked points, we have a number of admissible shapes,
namely all planted plane trees with valences $3,\ldots,k+1$ and with this number $m$ of 
labeled leaves. Such shapes obey the sum rule
\eqn\sumrule{m=1+\sum_{i=2}^k (i-1)p_i\ .}
We can then compute the relative weight $w(S)$ of a given shape $S$ among the admissible ones 
by considering all continuous history trees $H$ sharing this shape and integrating 
the distribution $\rho(H)$ over the lengths $x_j$, in number $n=1+\sum_{i=2}^k i p_i$.
We find
\eqn\wdesd{w(S)= \int_0^\infty \prod_{i=1}^n dx_i \ \rho(H)= \int_0^\infty dx {x^{n-1} \over (n-1)!}
\rho(H)\ ,}
where the variable $x$ stands for the total length of $H$.
This may be readily evaluated by multiplying both sides of eq.\invhist\ by $x^{n-1}/(n-1)!$ and
by integrating over $x$, leading to 
\eqn\invhistS{
{k^{{1\over k}} \sin\left({\pi \over k}\right)
\Gamma\left(1+{1\over k}\right) \over \pi}
\int_0^\infty {dM\over M} M^{m-{1\over k}} e^{-{M\over k}} 
\ w(S)=  \prod_{i=2}^k (\mu_i)^{p_i} }
where the $n$-dependent terms have cancelled out, henceforth
\eqn\wdes{w(S) =
{(-1)^{m-1}\over k^m} {\Gamma\left({1\over k}+1-m\right)\over \Gamma\left({1\over k}+1\right)}
\prod\limits_{i=2}^k (\mu_i)^{p_i} \ .}
Note that, as the number of leaves $m$ is fixed, we may rewrite
\eqn\wdeSun{w(S)= {1\over z_m} \prod_{i=2}^k (\mu_i)^{p_i}\ ,} 
displaying the universal weight of a shape in the MCRT${}_k$ as the product of the universal weights
$\mu_i$ of eq.\valmui\ at its branching points. The global prefactor
\eqn\globzm{z_m=(-1)^{m-1}k^m
{ \Gamma\left({1\over k}+1\right) \over \Gamma\left({1\over k}+1-m\right)
}}
ensures the proper normalization $\sum_S w(S)=1$, where the sum runs over
all shapes with $m$ leaves. This may be checked as follows: let $Y$ denote 
the (grand-canonical) generating function for {\it unmarked} shapes, i.e. shapes
whose leaves are not labeled, and with weights $\mu_i$ per $(i+1)$-valent branching 
point ($i=2,\ldots,k$) and $\mu_0$ per leaf. We have the relation
\eqn\eqforY{Y=\mu_0+\sum_{i=2}^k \mu_i Y^i=\mu_0+{k Y-1+(1-Y)^k\over k}\ ,}
where we used the explicit values \valmui\ for $\mu_i$ with $i\geq 2$. 
This relation may be inverted into 
\eqn\Ytomu{Y=1-(1-k\mu_0)^{1\over k}\ ,}
leading to the desired sum rule
\eqn\ruleshapes{\sum_{S  {\rm \ with} \atop m {\rm \ leaves}} \prod_{i=2}^k (\mu_i)^{p_i(S)}= 
m!\ Y\vert_{\mu_0^m} = (-1)^{m-1}k^m 
{ \Gamma\left({1\over k}+1\right) \over \Gamma\left({1\over k}+1-m\right)}=z_m\ , 
}
where the factor $m!$ accounts for the possible labelings of the $m$ leaves. 
\fig{The $11$ possible shapes with $m=4$ leaves when $k=4$. Each shape should
be supplemented by a labeling of its leaves, with $4!$ possible labelings per shape.
The universal weight of a shape is the product of the $\mu_i$ factors at its branching points, 
as displayed, times a global factor $1/z_4$. Here we have $z_4=231$, as given by eq.\globzm\ for
$k=m=4$, while $\mu_2=3/2$, $\mu_3=-1$ and $\mu_4=1/4$ from eq.\valmui. We easily
check the sum rule $(\mu_4 + 5 \mu_2\mu_3 + 5 \mu_2^3)\times (4!/z_4)=1$, as expected.}{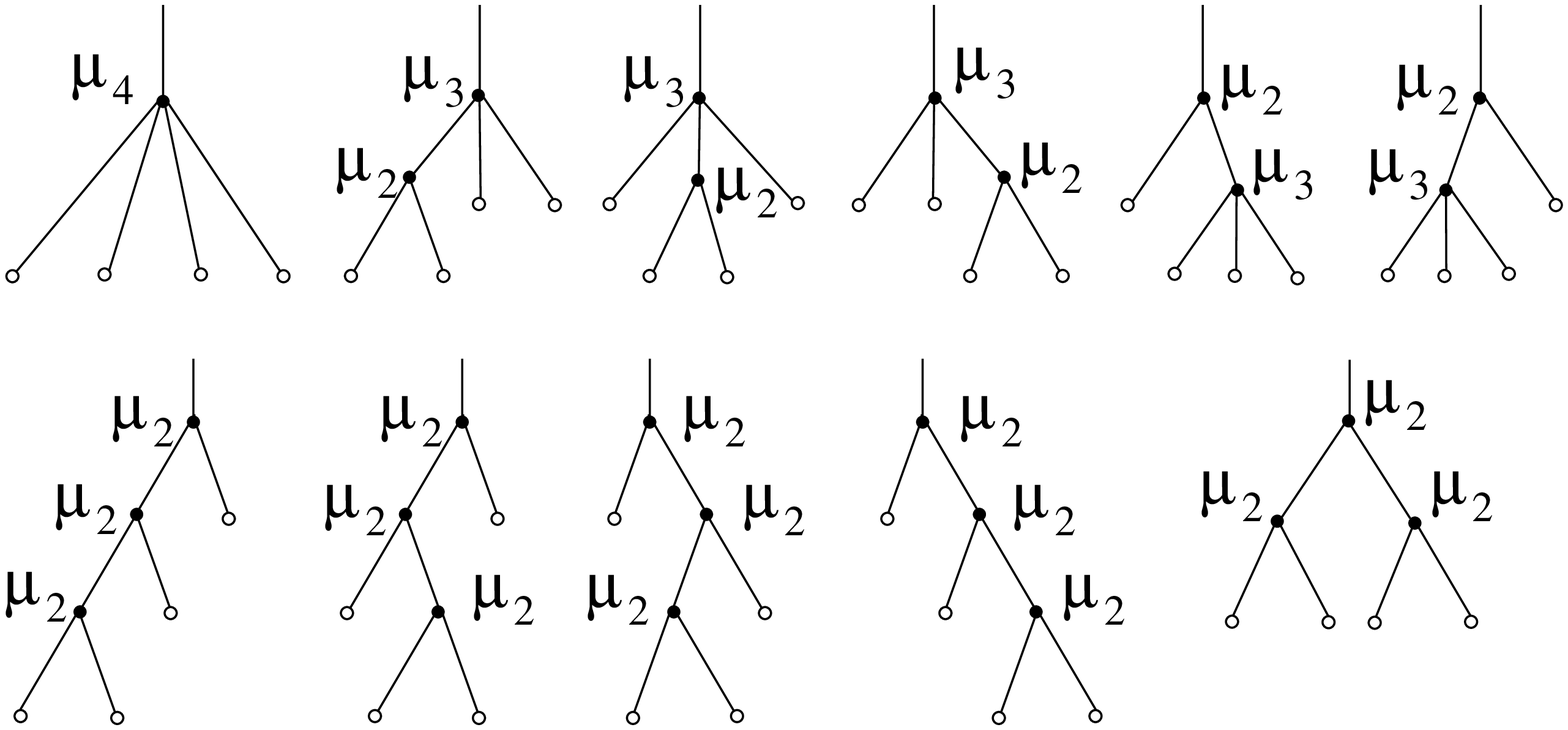}{11.cm}
\figlabel\shapes
\noindent For illustration, we have represented in fig.\shapes\ the list of all possible shapes 
with $m=4$ when $k=4$, together with the associated universal weights. 
Note that for $k=2$, as $p_2=m-1$ is entirely fixed, the law \wdeSun\ is uniform
over all admissible shapes, in agreement with eq.\probasha.

\newsec{MCRT: direct universal formulation} 

\subsec{Axiomatic approach}

In this section, we propose an alternative derivation of the MCRT${}_k$ average profile and history
distributions directly in a continuous formulation, without having recourse to discrete models, 
but with a minimal number of reasonable hypotheses. The MCRT${}_k$ describes a (possibly signed) measure 
over an ensemble of abstract trees with a {\it fixed total size}, say $1$, and we can
characterize its properties by the induced measure on continuous marked history trees
obtained by first picking leaves independently and uniformly at random in the abstract trees and
then extracting the subtree spanned by the root and these (marked) leaves.
The MCRT${}_k$ will be constructed so that it allows for histories with up to $(k+1)$-valent 
branching points while histories including $(k+2)$-valent branching points and more 
have a vanishing measure. 

Here we wish to define directly this measure, still denoted $\rho(H)$, where continuous marked 
history trees $H$ are as before characterized by a shape $S$ (taken to be
a planted plane tree with labeled leaves and with numbers $p_i$ of
$(i+1)$-valent branching points) endowed with branch lengths $x_j$. 

As just mentioned, the first requirement for the MCRT${}_k$ is that the only possible valences
for branching points in history trees belong to $\{3,4, \ldots, k+1\}$.
\fig{A example of rearrangement of a continuous tree, obtained by (a) placing
oriented markers on the branches of the tree, (b) cutting at these markers, (c) exchanging
the obtained pieces and finally (d) re-gluing them so as to built a new tree. Note that the
lengths of the various segments of branches are preserved at each step of the
rearrangement.}{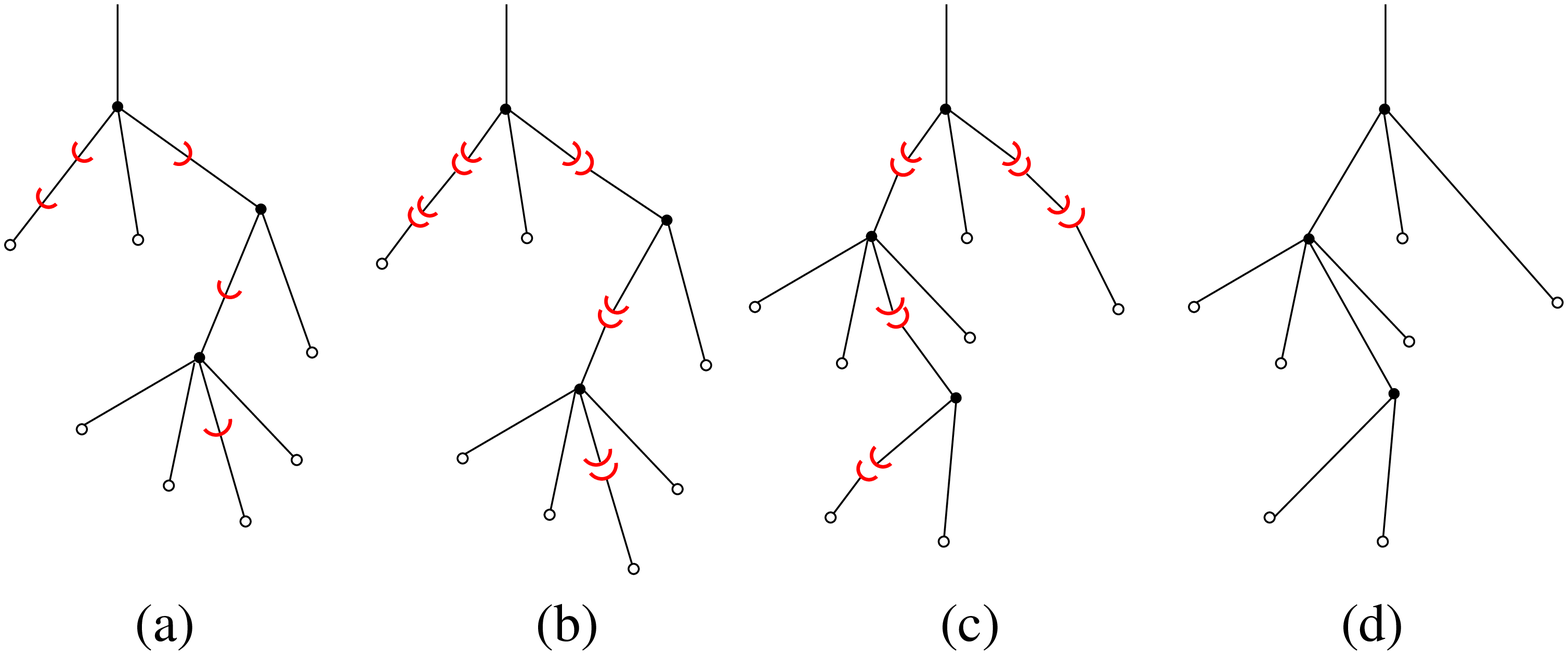}{12.5cm}
\figlabel\rearrange
Secondly, we assume that the measure is invariant under {\it rearrangements} of the
trees, inspired by a similar property for discrete trees, and defined as follows. 
For a given continuous history tree $H$, we may distribute a number 
of (oriented) markers on its branches and then cut the tree at these markers (see
fig.\rearrange\ for an illustration). The resulting pieces
are then rearranged in such a way as to preserve the orientation of markers and 
to produce another connected history tree $H'$. The above invariance means that
$\rho(H)=\rho(H')$. The consequences of this invariance are twofold. For a fixed shape, 
the measure of histories sharing that shape depends only on the {\it total} length 
$x\equiv \sum_j x_j$ of $H$. This simply uses the possibility within the above 
rearrangements to redistribute, at fixed shape, internal segments within branches.
Moreover, the dependence on the shape is only through its numbers $p_i$ of $(i+1)$-valent 
branching points. This uses the possibility, as part of the above rearrangements, 
of permuting the branching points by appropriate cutting of branches. 
With the above hypotheses, we may write $\rho(H)$ in the form
\eqn\rhophi{\rho(H)=\phi(x;\{p_2,p_3,\ldots,p_k\})}
for all $H$ with fixed values of $p_i$, $i=2,\ldots,k$, and fixed total length $x$.
\fig{The two possible ways of adding an extra leaf to a given history 
(see text).}{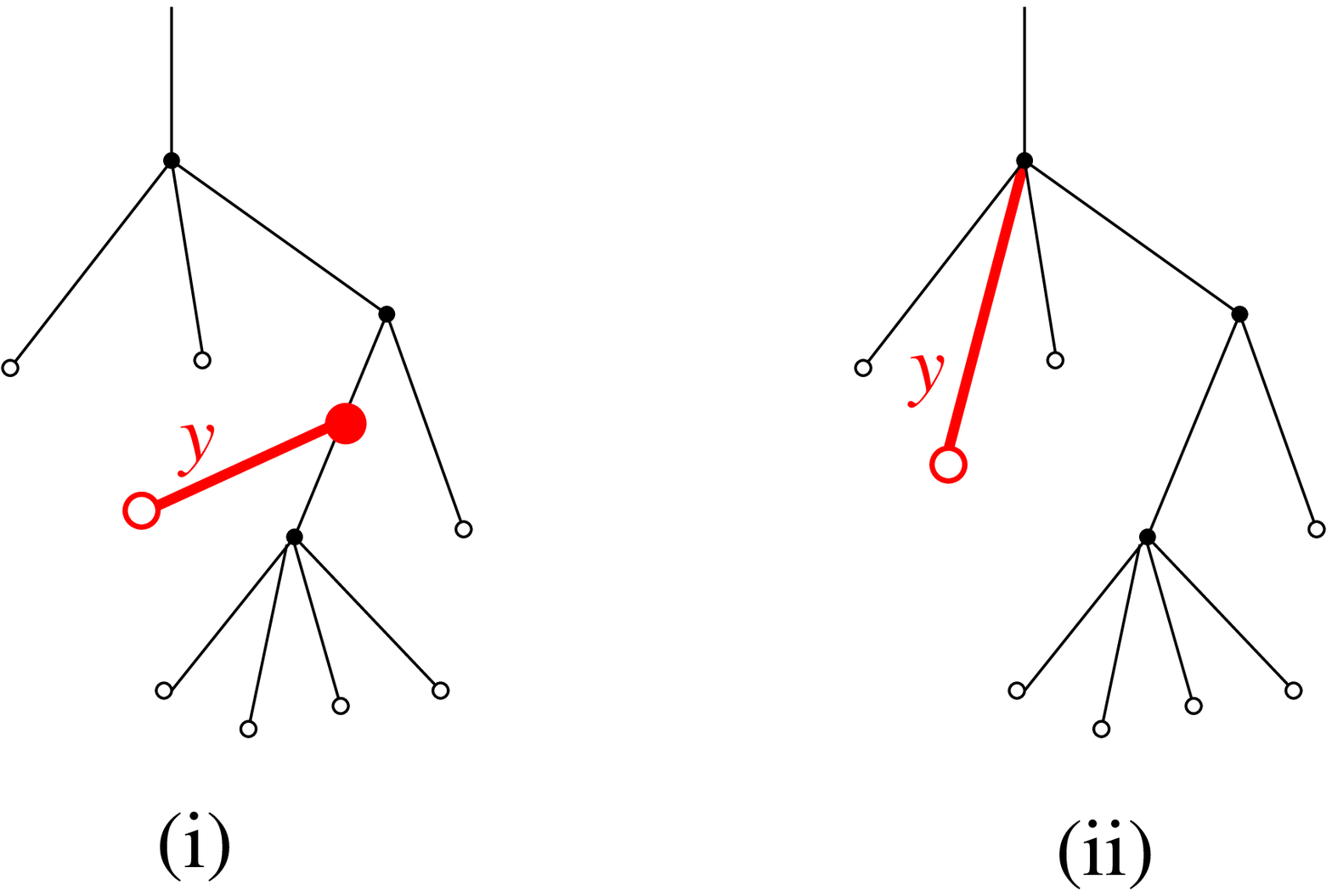}{8.cm}
\figlabel\consistency
We may now write an infinite set of consistency equations relating the $\phi$'s at different
values of their arguments as follows. Given a history $H$ with $m$ leaves, we may
consider all histories $H'$ obtained by adding to $H$ one extra leaf and express
that $\rho(H)$ is the integrated measure over these histories $H'$. More precisely,
as illustrated in fig.\consistency, there are two ways of adding an extra leaf: 
\item{(i)} by creating a new $3$-valent branching
point along an existing branch of $H$, splitting that branch into two pieces and
creating a new branch, say of length $y$, connected to the extra leaf. Integrating
over the position of this new branching point produces a factor $2x$ equal to the external
perimeter of $H$. Note that in $H'$, the numbers of branching points read $p'_i=p_i$ 
for $i>2$ and $p'_2=p_2+1$.
\item{(ii)} by connecting the extra leaf to an existing $(j+1)$-valent branching point (with
$j<k$) via a new branch of length $y$,
thus replacing it by a $(j+2)$-valent one. As we are dealing with plane trees, there
are $j+1$ inequivalent ways of doing so. In $H'$, the numbers of branching points read
$p'_i=p_i$ for $i\neq j,j+1$, while $p'_j=p_j-1$ and $p'_{j+1}=p_{j+1}+1$.
\par
\noindent Integrating over the length of the extra branch $y$ and using the form \rhophi\ for
$\rho(H)$, we deduce the consistency relations
\eqn\consist{\eqalign{\phi(x;& \{p_2,\ldots,p_k\}) = 2 x \int_0^\infty dy \ \phi(x+y;\{p_2+1,p_3,\ldots,p_k\})
\cr & +\sum_{j=2}^{k-1} (j+1)p_j \int_0^\infty dy \ \phi(x+y;\{p_2,\ldots,p_{j-1},p_j-1,p_{j+1}+1,p_{j+2},
\ldots,p_k\})\ .\cr}}
Note that, when all $p_i$'s are equal to $0$, $\phi$ is nothing but the average profile,
namely:
\eqn\prodi{\rho(x)=\phi(x;\{0,\ldots,0\})\ .} 
This function is assumed to be sufficiently regular, and {\it exponentially decreasing} at
large $x$ so that, for instance, all its moments are finite.

To proceed further, we rely on a {\it factorization assumption} and write all $\phi$'s as
\eqn\locasump{\phi(x;\{p_2,\ldots,p_k\})= \prod_{i=2}^k \left( O_i \right)^{p_i}\ \rho(x)\ ,}
stating that each $(i+1)$-valent branching point contributes via the action of an operator $O_i$
on $\rho(x)$, where these operators commute for different indices. 
The relation \locasump\ should be universal in the sense that it should remains the same 
for all ensembles $E_M$ of abstract trees with fixed total size $M$ rather than $1$.
For trees in $E_M$, the history distributions and average profile are rescaled into
\eqn\phiM{\phi_M(x,\{p_2,\ldots,p_k\})={M^m \over M^{n\,\nu_k}}\phi\left({x\over M^{\nu_k}};
\{p_2,\ldots,p_k\}\right),\qquad \rho_M(x)={M \over M^{\nu_k}}\rho\left({x\over M^{\nu_k}}\right)} 
where $n=1+\sum_{i=2}^k p_i$ is the number of branches, $m=1+\sum_{i=2}^k (i-1) p_i$ the
number of marked leaves, and where $\nu_k$ is the inverse of the fractal dimension 
$d_k$ of the MCRT, yet to be computed. 
To ensure that \locasump\ still holds with the substitutions $\phi\to\phi_M$ and $\rho\to\rho_M$, 
we need that the operator $O_i$ have a non trivial {\it scaling dimension} $\alpha_i$, namely satisfy
\eqn\scalO{g(x)=f(\Lambda x) \Rightarrow \big[O_i g\big](x) = \Lambda^{\alpha_i} \big[O_i f\big](\Lambda x)}
for some real $\alpha_i$.
Rewriting eq.\locasump\ for the $\phi_M$'s, we also get the consistency relation 
\eqn\consirel{m-n \nu_k=1-\nu_k\left(1+\sum_{i=2}^k \alpha_i p_i\right)}
which must hold for all values of the $p_i$'s. This yields the scaling dimensions
\eqn\valalph{\alpha_i=1+{\nu_k-1\over \nu_k} (i-1)\ .}

We now look for a solution to the consistency relations \consist\ in the form \locasump,
and make a simple educated guess on the operators $O_i$, namely that: 
\eqn\ansatz{O_i= \mu_i\, (-d)^{\alpha_i}}
involving a scalar multiplicative factor $\mu_i$ and a power $\alpha_i$ of (minus) the derivative
operator $(-d)$ with respect to $x$, with $\alpha_i$ as in eq.\valalph. 
At this stage, this particular form is only an Ansatz inspired
from the discrete approach of section 2 and which,
as we shall see, turns out to yield a fully consistent solution to eq.\consist\ for a specific
value of the scaling exponent $\nu_k$ in eq.\valalph. We believe that this solution is 
actually unique provided all $\phi$'s are exponentially decreasing at large $x$. 

It is known that there are various (inequivalent) 
definitions of the pseudo derivative operators $(-d)^\alpha$ but here, we are led 
to a very natural choice by noting that we let them act on a function $\rho(x)$  such
that all its derivatives vanish when $x\to \infty$. We define the fractional integrals
and derivatives as:
\eqn\defdalp{\eqalign{(-d)^{-\beta} f (x)& \equiv {1\over \Gamma(\beta)} \int_0^\infty dt\ t^{\beta-1} 
e^{t\, d} f(x)\cr 
&= {1\over \Gamma(\beta)} \int_x^\infty du (u-x)^{\beta-1} f(u)\cr}}
for positive $\beta$, and, for positive $\alpha$, 
\eqn\defdalpos{(-d)^\alpha= (-d)^{[\alpha]+1} (-d)^{-\beta} \quad {\rm with} 
\quad \beta\equiv[\alpha]+1-\alpha}
where $[\alpha]$ stands for the integer part of $\alpha$, so that $\beta$ lies is $(0,1]$.
In eq.\defdalp,  we have used the standard shift operator $e^{t\, d}f(x)=f(x+t)$. The above
definitions coincide with the so-called Weyl fractional integral \HILF.

Apart from the desired scaling property \scalO\ easily checked from our definition 
\defdalpos\ with $\alpha=\alpha_i$, the pseudo derivative operators obey the additivity property and 
generalized Leibniz formula
\eqn\addiLeib{(-d)^\alpha (-d)^\beta = (-d)^{\alpha +\beta}\quad,\quad (-d)^\alpha x = x(-d)^\alpha
-\alpha (-d)^{\alpha-1}}
when acting on functions which are exponentially decreasing at infinity. Note that the 
derivation of these properties involve integrations by parts and it is crucial that all
boundary terms at infinity vanish, as a consequence of the exponential decrease. 
We shall also use
\eqn\dmoinsun{(-d)^{-1}\phi(x)=\int_0^\infty dy\ \phi(x+y)} 
to replace the integrals in eq.\consist\ with pseudo derivative operators.
As we shall see below, the values of $\alpha_i$ will be fixed to be rationals of the form $r/(k-1)$, 
hence the reader may alternatively think
of $\rho(x)$ as having an integral representation of the form \intphi\ with some ${\tilde\rho}(\xi)$
to be determined, in which case the action of the pseudo derivative operators above 
is equivalently defined via eq.\resintphi.

The solution is further fixed by substituting the form \locasump\ with $O_i$ as in
\ansatz\ in the consistency relations \consist. The latter now read
\eqn\consisD{{\cal D} \rho =2 x \mu_2 {\cal D}(-d)^{\alpha_2-1} \rho +\sum_{j=2}^{k-1}
(j+1) p_j {\mu_{j+1}\over \mu_j} {\cal D} (-d)^{\alpha_{j+1}-\alpha_j-1} \rho}
where ${\cal D}=(-d)^{\sum_i \alpha_i p_i}$. Using the Leibniz formula in \addiLeib\ to
commute $x$ through the operator ${\cal D}$, we deduce
\eqn\commucalD{{\cal D}\!\left\{\!\!\Big(\rho\!-\! 2 x \mu_2 (-d)^{\alpha_2-1}\rho \Big)\!\!
-\!\!\left(\sum_{j=2}^{k-1} p_j\!\left( 2 \mu_2 \alpha_j\! +\! (j\!+\!1) {\mu_{j+1}\over \mu_j}\! \right)\! 
+ 2 \mu_2 \alpha_k p_k \right)\!(-d)^{\alpha_2-2} \rho \!\right\}=0}
where we have used $\alpha_{j+1}-\alpha_j-1=\alpha_2-2$ from eq.\valalph.
This must hold for all values of the $p_i$'s whose coefficients must therefore all
vanish. The $p_k$ term yields $\alpha_k=0$, from which we deduce
\eqn\valnuk{\nu_k={k-1\over k}}
and the scaling dimensions 
\eqn\alphval{\alpha_i={k-i\over k-1}\ .}
The $p_j$ terms for $j=2,\ldots,k-1$ fix the relations
\eqn\relmj{{\mu_{j+1}\over \mu_j}=-{2\mu_2 \over j+1}{k-j\over k-1}}
hence
\eqn\valmuj{\mu_j\equiv{(-1)^j \over k} {k \choose j}\ \left({2 \mu_2 \over k-1}\right)^{j-1}\ .}
Finally, we are left with a set of relations for $\rho$ valid for all $p_i$'s and implied by
the unique relation obtained when all $p_i$'s are equal  to $0$, in which case ${\cal D}=1$:
\eqn\fracdifbis{\rho(x) = 2\mu_2 x \, (-d)^{-{1\over k-1}} \rho(x) \ . }
The factor $\mu_2$ in eqs.\valmuj\ and \fracdifbis\ is arbitrary and may be reabsorbed into
a rescaling of the length $x \to \Lambda x$ . We decide to fix $\mu_2= (k-1)/2$, in which case
eq.\valmuj\ matches eq.\valmui\ of the discrete approach and we end up with the fractional
differential equation for the average profile $\rho(x)$:
\eqn\fracdif{\eqalign{\rho(x)& = (k-1) x \, (-d)^{-{1\over k-1}} \rho(x) \cr
& ={(k-1)x \over \Gamma\left({1\over k-1}\right)}
\int_x^\infty du (u-x)^{{1\over k-1}-1 }\rho(u) \ .\cr}}
As this stage, we note that if we interpret the fractional derivative according
to eqs.\intphi\ and \resintphi, we immediately see that the solution $\rho(x)$ given 
by \intformu\ in the discrete approach does indeed satisfy the equation \fracdif. 
As we shall see below, we need not to have recourse to any integral representation
to prove that eq.\fracdif\ has a unique solution with exponential decay at large $x$,
and that moreover, this solution matches that of the discrete approach.
This is the aim of the next three sections.

\subsec{Ordinary differential equation for $\rho(x)$}

In this section, we derive an {\it ordinary} differential equation for
the average profile $\rho(x)$ based on manipulations of fractional derivatives
as follows. We start from the fractional differential equation in the first line of 
\fracdif\ and apply the operator $-(-d)^{k\over k-1}$ on both sides. 
Using the generalized Leibniz formula \addiLeib\ for $\alpha=k/(k-1)$, we get the relation
\eqn\fondafracdif{-(-d)^{{k\over k-1}}\rho = (k-1) x \rho' + k \rho} 
which, upon introducing the operators 
\eqn\ops{\delta \equiv -(-d)^{k\over k-1}\quad {\rm and} \quad P\equiv (k-1) x d +k\ , } 
may be recast into
\eqn\opfracdif{\delta \rho = P \rho \ .}
The two operators $\delta$ and $P$ obey the following commutation relation:
\eqn\opreltwo{[\delta,P]= k \delta }
which is a direct consequence of the generalized Leibniz formula
\addiLeib\ for $\alpha=k/(k-1)$.
We may then show by induction that, for all $j=1,2,\ldots $:
\eqn\deltajrho{\delta^j \rho= (P+(j-1) k)(P+(j-2) k)\cdots (P+k)P \rho\ .}
This relation holds clearly for $j=1$ from eq.\opfracdif\ and is simply proved by recursion 
upon noticing that, from eq.\opreltwo,  $\delta (P+a)= (P+a+k) \delta$ for any constant $a$. 
Taking $j=k-1$ and using $\delta^{k-1}=-\,d^k $, this finally
yields an ordinary differential equation of order $k$ for $\rho$, namely
\eqn\ordif{-d^k \rho = \prod_{j=0}^{k-2} (P+k j) \rho\ .}
For illustration, when $k=2,3,4$, these equations read respectively:
\eqn\firstordif{\eqalign{
0& = 2 \rho + x \rho' +\rho'' \cr
0& = 18 \rho + 22 x \rho' + 4 x^2 \rho'' +\rho''' \cr
0& = 384 \rho +771 x \rho' + 297 x^2 \rho'' + 27 x^3 \rho'''  + \rho^{(4)}\cr}}

\subsec{Explicit solution} 

We will now determine the function $\rho(x)$ as the suitably normalized solution 
of eq.\ordif\ with exponential decay at large $x$. As we shall see, this latter condition 
turns out to be sufficient to select a unique solution.

Let us first look for the general solution of eq.\ordif\ in the form of a series expansion
\eqn\devrho{\rho(x) = \sum_{m\geq 0} \rho_m x^m\ .}
Eq. \ordif\ translates into recursion relations on the coefficients, namely
\eqn\recurcoeff{{\rho_{m+k}}= - {m!\over (m+k)!} \rho_m \prod_{j=0}^{k-2} (m (k-1)+k + k j)}
which determines all coefficients in term of the first $k$ ones $\rho_0, \rho_1, \cdots \rho_{k-1}$.
We immediately get the general solution:
\eqn\solrho{\rho(x)=\sum_{p=0}^{k-1} \rho_p x^p \quad {}_{k-1}\!F_{k-1}\left(\{ a_i(p)\},
\{b_i(p)\};
-{(k-1)^{k-1}\over k} x^k\right)}
in terms of $k$ generalized hypergeometric series ${}_{k-1}\!F_{k-1}$ with arguments 
\eqn\argu{\eqalign{ a_i(p)& =\left\{\matrix{ \displaystyle{{p\over k}+{i\over k-1}} 
& \quad {\rm for}\quad i=1,2,\ldots, k-2 
\cr & \cr  1 & \quad {\rm for} \quad i=k-1 \hfill \cr} \right. \cr
b_i(p)& ={p\over k}+{i\over k} \quad {\rm for}\quad i=1,2,\ldots, k-1 \cr}}
for $p=0,1,\ldots,k-1$. Recall that the generalized hypergeometric series are defined via
\eqn\defhyper{{}_{p}\!F_{q}\left(\{ a_i\},
\{b_j\}; z \right)\equiv
\sum_{m\geq 0} {\prod\limits_{i=1}^{p}(a_i)_m \over \prod\limits_{j=1}^q (b_j)_m} {z^m\over m!}
}
with $(a)_m\equiv a(a+1)\cdots(a+m-1)$.
For simplicity, we rewrite the general solution \solrho\ as the linear combination
\eqn\resolrho{\rho(x)=\sum_{p=0}^{k-1} C_p F_p(x)}
with $k$ new arbitrary coefficients $C_p$, and where we define the suitably normalized
functions:
\eqn\fp{\eqalign{& F_p(x)\equiv  \left({(k-1)^{k-1}\over k}\right)^{{p\over k}} x^p 
\prod_{i=1}^{k-1}{\Gamma(a_i(p))\over \Gamma(b_i(p))} \ 
{}_{k-1}\!F_{k-1}\left(\{ a_i(p)\},
\{b_i(p)\};
-{(k-1)^{k-1}\over k} x^k\right)\cr
& = \sqrt{{k\over 2 \pi (k-1)}} k^{{k-1\over k}p}\Gamma\left({k-1\over k}p+1\right)\ {x^p\over p!}\ 
{}_{k-1}\!F_{k-1}\left(\{ a_i(p)\},
\{b_i(p)\};
-{(k-1)^{k-1}\over k} x^k\right).
\cr}}
Here we have used the simplification
\eqn\superprod{\prod_{i=1}^{k-1}{\Gamma(a_i(p))\over \Gamma(b_i(p))}={\prod\limits_{i=1}^{k-2}\Gamma\left(
{p\over k}+{i\over k-1}\right) \over \prod\limits_{i=1}^{k-1}\Gamma\left( {p\over k}+{i\over k}\right)}
=\sqrt{{k\over 2\pi (k-1)}} {k^p\over (k-1)^{{k-1\over k}p}}
{\Gamma\left(
{k-1\over k}p+1\right) \over \Gamma( p+1)}} 
obtained as a consequence of the Gamma product theorem, stating that:
\eqn\gammprod{\Gamma(n x)= (2\pi)^{{1-n\over 2}}n^{n x-{1\over 2}}
\prod_{i=0}^{n-1}\Gamma\left(x+{i\over n}\right)\ .}
Remarkably, the coefficients $C_p$ may be completely fixed by simply requiring 
that 
\item{(i)} $\rho(x)$ has an exponential decay for large (positive) $x$, and 
\item{(ii)} $\rho(x)$ is normalized.
\par
To implement condition (i) above, we use the following asymptotics \WOLF\ for hypergeometric 
series of argument $(-z)$ at large positive values of $z$: 
\eqn\asympt{\eqalign{
\prod_{i=1}^{k-1}{\Gamma(a_i)\over \Gamma(b_i)}& \quad
{}_{k-1}\!F_{k-1}\left(\{ a_i\}, \{b_i\};
-z\right)\cr  = & \sum_{j=1}^{k-1} {\Gamma(a_j)\over z^{a_j}} {\prod\limits_{\ell\neq j}\Gamma(a_\ell-a_j)
\over \prod\limits_\ell \Gamma(b_\ell-a_j)} \quad {}_k\!F_{k-2}\left(\{a^{(j)}_i\},\{b^{(j)}_i\};
{1\over z}\right)+{\cal O}(z^\chi e^{-z})\cr}}
where $\chi=\sum_i(a_i-b_i)$ and with
\eqn\newargu{\eqalign{ a^{(j)}_i& =\left\{\matrix{ 
a_j-b_i +1  & \quad {\rm for} \quad i=1,2,\ldots, k-1 \cr
a_j & \quad {\rm for}\quad i=k \hfill \cr 
} \right. \cr
b^{(j)}_i& = \left\{\matrix{ a_j-a_i +1 \hfill  &  \quad {\rm for}\quad i=1,2,\ldots, j-1 \hfill \cr 
a_j-a_{i+1}+1 & \quad {\rm for} \quad i=j,j+1,\ldots k-2 \ .\cr } \right. \cr}}
These asymptotics may be applied to the list of arguments $\{a_i=a_i(p)\}$ and $\{b_i=b_i(p)\}$
of eq.\argu\ with $p=0,1,\ldots k-1$, giving rise to sets  $\{a^{(j)}_i(p)\}$ and $\{b^{(j)}_i(p)\}$.
First, we notice that $a^{(j)}_k(p)=b^{(j)}_{k-2}(p)=a_j(p)$, hence their respective contributions
to the ${}_k\!F_{k-2}$ hypergeometric series cancel out. 
Moreover, for $j\leq k-2$, the remaining arguments $\{a^{(j)}_i(p)\}_{i\leq k-1}$ and 
$\{b^{(j)}_i(p)\}_{i\leq k-3}$ all turn out to be {\it independent} of $p$,
hence the functions
\eqn\bj{B_j(z)\equiv \quad {}_k\!F_{k-2}\left(\{a^{(j)}_i(p)\},\{b^{(j)}_i(p)\};
{1\over z}\right)}
for $j=1,\ldots k-2$ are independent of $p$. Concerning the $j=k-1$ term in \asympt, 
we note that, as soon as $p\geq 1$,  as both $a_{k-1}(p)$ and $b_{k-p}(p)$ are equal to $1$, 
the corresponding prefactor in eq.\asympt\ vanishes identically. The $j=k-1$ term only contributes 
when $p=0$ and involves a function:
\eqn\bjzero{B_{k-1}(z)\equiv \quad {}_k\!F_{k-2}\left(\{a^{(k-1)}_i(0)\},\{b^{(k-1)}_i(0)\};
{1\over z}\right)\ .}

The large $x$ asymptotics of $\rho$ read therefore 
\eqn\asymprhobis{\eqalign{
\rho(x)&=
\sum_{j=1}^{k-2} B_j(z) \sum_{p=0}^{k-1} C_p 
z^{{p\over k}-a_j(p)}
\Gamma(a_j(p))
{\prod\limits_{\ell\neq j}\Gamma(a_\ell(p)-a_j(p))
 \over \prod\limits_\ell \Gamma(b_\ell(p)-a_j(p))}
\cr
&\ \ \ \ + B_{k-1}(z) C_0 z^{-1}
 {\prod\limits_{\ell\neq k-1}\Gamma(a_\ell(0)-1)
  \over \prod\limits_\ell \Gamma(b_\ell(0)-1)}+{\cal O}(z^{1\over 2} e^{-z})
\cr 
&= \sum_{j=1}^{k-2}B_j(z) z^{-{j\over k-1}}{\prod\limits_{\ell=1 \atop \ell\neq j}^{k-2}\Gamma\left(
{\ell\over k-1}-{j \over k-1}\right)\over \prod\limits_{\ell=1}^{k-1} \Gamma\left(
{\ell\over k}-{j\over k-1}\right)} \left\{\sum_{p=0}^{k-1}C_p \Gamma(a_j(p))\Gamma(1-a_j(p))
\right\}
\cr
&\ \ \ \ + B_{k-1}(z) z^{-1} {\prod\limits_{\ell=1 }^{k-2}\Gamma\left(
{\ell\over k-1}-1\right)\over \prod\limits_{\ell=1}^{k-1} \Gamma\left(
{\ell\over k}-1\right)} C_0+{\cal O}(z^{1\over 2} e^{-z})
\cr
}
}
with $z=x^k (k-1)^{k-1}/k$. Note that, to get the precise form of the exponentially decreasing
correction, we have made use of the relation $p/k+\chi=p/k+\sum(a_i(p)-b_i(p))=1/2$ irrespectively of $p$.

Requiring an exponential decay for $\rho(x)$ at large $x$ imposes that $C_0=0$ and
\eqn\eqforcoef{\sum_{p=1}^{k-1}{C_p \over \sin\left(\pi\left({p\over k}+{j\over k-1}\right)\right)}=0}
for $j=1,\ldots,k-2$ by use of the identity $\Gamma(x)\Gamma(1-x)=\pi/\sin(\pi x)$.
This fixes all $C_p$ for $p=1,\ldots k-1$ up to a global factor $C$, namely
\eqn\valcp{C_p=C \sin\left(\pi {p\over  k}\right)\ .}
This may be checked by use of the identity
\eqn\eqforcoef{\sum_{p=1}^{k-1}{\sin\left(\pi {p\over  k}\right)  \over 
\sin\left(\pi{p\over k}+\alpha \right)} = k{\sin ((k-1) \alpha)\over \sin(k \alpha)}}
which precisely vanishes at $\alpha=\pi j/(k-1)$ for $j=1,2,\ldots,k-2$, as wanted.
Picking the values \valcp\ for $C_p$, we are left with the leading behavior 
$\rho(x)\propto z^{1\over 2} e^{-z})$, in agreement with eq.\asymprho. 

Finally the remaining constant $C$ is fixed by the normalization condition (ii) above. 
This is best seen by integrating the differential equation \ordif\ on $[0,\infty )$, leading to
\eqn\rkzero{\rho^{(k-1)}(0)
=\prod_{j=0}^{k-2}(1+k j)
\int_0^\infty \rho(x)dx
=\prod_{j=0}^{k-2}(1+k j)\ ,}
where we have used the fact that $\int_0^\infty P f =\int_0^\infty f$ for any exponentially 
decaying differentiable function $f$. This is to be compared with
\eqn\rhokzero{\eqalign{\rho^{(k-1)}(0)& = C_{k-1}\ F_{k-1}^{(k-1)}(0)\cr 
&= C \sin\left( \pi {k-1\over k}\right)
\sqrt{{k\over 2 \pi (k-1)}}\ k^{{1\over k}+k-2}\ \Gamma\left({1\over k}+k-1\right)
\cr}}
obtained from eqs.\resolrho\ and \fp\ where only the $p=k-1$ term contributes. 
This immediately leads to
\eqn\valC{\eqalign{C = \sqrt{{2 \pi (k-1) \over k}} {1\over k^{1\over k} \Gamma\left({1\over k}+1\right)
\sin\left({\pi\over k}\right) }\cr}
}
and, upon substitution into eq.\resolrho\ of the values of the coefficients $C_p$ read off 
eqs.\valcp, we get the final explicit formula:
\eqn\formurho{\rho(x)=\sum_{p=1}^{k-1} {\sin\left(\pi {p\over k}\right)\over \sin\left({\pi \over k}\right)}
{k^{{k-1\over k}p}\Gamma\left({k-1\over k}p+1\right)\over k^{1\over k}\Gamma\left({1\over k}+1\right)}
{x^p\over p!}\ {}_{k-1}\!F_{k-1}\left(\{a_i(p)\},\{b_i(p)\};-{(k-1)^{k-1}\over k}x^k \right)}
where the ${}_{k-1}\!F_{k-1}$ functions in practice reduce to ${}_{k-2}\!F_{k-2}$ functions as
$a_{k-1}(p)=b_{k-p}(p)=1$ for all $p=1,\ldots,k-1$. 
For illustration, when $k=2,3,4$ this yields:
\eqn\explitrho{\eqalign{
k=2: \quad \rho(x) &= x e^{-{x^2\over 2}}\cr
k=3: \quad \rho(x) &=
3^{1\over 3} x {\Gamma\left({5\over 3}\right)\over \Gamma\left({4\over 3}\right)}
\ {}_1\!F_{1}\left({5\over 6},{2\over 3}, -{4\over 3}x^3 \right)
+2 x^2 \ {}_1\!F_{1}\left({7\over 6},{4\over 3}, -{4\over 3}x^3 \right)
\cr
 k=4: \quad \rho(x) &= 
2 x {\Gamma\left({7\over 4}\right)\over \Gamma\left({5\over 4}\right)}
\ {}_2\!F_{2}\left(\left\{{7\over 12},{11\over 12}\right\},\left\{{1\over 2},{3\over 4}\right\}
, -{27\over 4}x^4 \right)\cr
&\  +
4 x^2 {\Gamma\left({5\over 2}\right)\over \Gamma\left({5\over 4}\right)}
\ {}_2\!F_{2}\left(\left\{{5\over 6},{7\over 6}\right\},\left\{{3\over 4},{5\over 4}\right\}
, -{27\over 4}x^4 \right)\cr
&\  +
{15\over 2} x^3
\ {}_2\!F_{2}\left(\left\{{13\over 12},{17\over 12}\right\},\left\{{5\over 4},{3\over 2}\right\}
, -{27\over 4}x^4 \right)\ .
\cr
}}
For $k=2$, we recover the expression \rhoCRT, as expected.  
As we shall see in the next section, the above expressions match the integral formulas used
for the plot of fig.\rhoplot.

\subsec{Compatibility with the integral formulas derived from discrete models}
Let us now compare the result \formurho\ to the integral formula \intformu\ of sect.2.3.
Expanding this integral formula in $x$, we use 
\eqn\expim{\eqalign{{\rm Im}\left\{e^{\tau \xi^{k-1} x}\right\}& =\sum_{\ell=0}^\infty \sin\left( 
\pi {\ell \over k}\right) \xi^{(k-1)\ell} {x^\ell\over \ell !} \cr &
=\sum_{p=0}^{k-1} \sin\left(
\pi {p \over k}\right) x^p \sum_{m=0}^\infty (-1)^m \xi^{(k-1)(k m+p)} {x^{k m}\over (k m+p)! } \cr }}
to rewrite
\eqn\rerho{\rho(x)={1\over k^{1\over k} \Gamma\left({1\over k}+1\right)
\sin\left({\pi\over k}\right)}\sum_{p=1}^{k-1} \sin\left(
\pi {p \over k}\right) x^p A_p\left(-{(k-1)^{k-1}\over k}x^k \right)}
with 
\eqn\apz{\eqalign{A_p(z)& =\sum_{m=0}^\infty \left({k\over (k-1)^{k-1}}\right)^m {z^m\over (k m+p)!}
\int_0^\infty d\xi \xi^{k-1} e^{-{\xi^k\over k}} \xi^{(k-1)(k m+p)}\cr 
&= \sum_{m=0}^\infty {k^{k m +{k-1\over k}p}\over (k-1)^{(k-1)m}} {\Gamma\left(
(k-1)\left(m+{p\over k}\right)+1\right) \over \Gamma\left( k\left(m+{p\over k}\right)+1\right)} \ z^m\cr 
&= \sum_{m=0}^\infty \sqrt{{2\pi (k-1)\over k}} \left({(k-1)^{k-1}\over k}\right)^{p\over k}
{\prod\limits_{i=1}^{k-2}\Gamma\left(
m+{p\over k}+{i\over k-1}\right) \over \prod\limits_{i=1}^{k-1}\Gamma\left( m+{p\over k}+{i\over k}\right)}
\ z^m \cr }}
by use of the Gamma product theorem \gammprod.
Using
\eqn\gammarat{{\prod\limits_{i=1}^{k-2}\Gamma\left(
m+{p\over k}+{i\over k-1}\right) \over \prod\limits_{i=1}^{k-1}\Gamma\left( m+{p\over k}+{i\over k}\right)}
= {1\over m!}\prod_{i=1}^{k-1}
{\Gamma\left(a_i(p)\right) \over \Gamma\left( b_i(p)\right)}
\prod_{i=1}^{k-1}{(a_i(p))_m\over (b_i(p))_m} }
with $a_i(p)$ and $b_i(p)$ given by \argu, we identify
\eqn\AtoF{A_p(z)=\sqrt{{2\pi (k-1)\over k}} F_p(z)}
with $F_p$ as in eq.\fp. Eq.\rerho\ then boils down to $\rho(x)=\sum\limits_{p=1}^{k-1} C_p F_p(z)$
with $C_p$ as in \valcp\ and $C$ as is \valC. 
The two formulas \intformu\ and \formurho\ therefore {\it define the same function} $\rho(x)$, as wanted.
Finally, all history distributions of section 3.1, obtained by acting on $\rho(x)$  via eq.\locasump\ 
with the particular choices \ansatz\ and \alphval, do coincide with the integral expressions  
\rhohcont\ of section 2, as seen from eq.\rhohfrac. We conclude that the axiomatic approach developed
here fully agrees with the discrete results of section 2.

\newsec{Conclusion}

In this paper, we have introduced a generalization of Aldous' CRT describing the continuum
scaling limit of multicritical ensembles of discrete trees. Restricting to $k$-multicritical
ensembles leads to a unique universal limit, the MCRT${}_k$, characterized by continuous history
trees with branching points of valence $3$ to $k+1$. We have obtained the average profile 
and history distributions of the MCRT${}_k$ in the form of either integral representations
or combinations of hypergeometric series. The history distributions are expressed 
in terms of universal weights attached to the branching points, combined with fractional
derivative operators acting on the average profile. Except for the usual CRT ($k=2$),
the universal weights carry signs hence the measure on trees is not positive. This phenomenon
parallels a similar property for the scaling limits of 2DQG as obtained from a one-matrix
model \ONEMATH\ and is the signature of the non-unitary character of the underlying conformal 
field theory \CARDY.
Likewise, the solutions of 2DLG display the same alternating signs to ensure multicriticality \LORGRA.

Our approach was twofold: we first derived the MCRT profile and history distributions as scaling limits of 
explicit expressions obtained in discrete models, and then recovered the same results
from a purely continuous axiomatic approach, thereby corroborating the universal nature 
of the MCRT. A more abstract definition of the MCRT itself remains to be found. A possible
route would consist in using random walks as in the CRT. Indeed, from ref.\LORGRA, 
we already know that $k$-multicritical 2DLG can be rephrased in terms of random walks with 
properly weighted ascending steps of various lengths. Presumably, a continuous limit of
these walks will provide the generalization of Brownian motion needed for the definition
of the MCRT.

An important ingredient in our construction was the use of ensembles of trees with
fixed numbers of leaves. While in the Brownian CRT, the use of ensembles with
fixed numbers of edges turns out to lead to equivalent results, this is not the case
for the MCRT with $k>3$. For instance, we may consider trees with a fixed number
$N'$ of edges and with inner vertices weighted according to the minimal prescription \ming.
It is then easily seen that the partition function $Z_{N'}$ of this new ensemble has
the sign $(-1)^{N'-1}$ due to the fact that $g_i^*$ has the sign $(-1)^i$. Up to this global
sign, $Z_{N'}$ is identical to the partition function of the ensemble with $g_i=|g_i^*|$,
hence with only positive weights. The large $N'$ continuum limit is therefore governed
by the usual (non-multicritical) Brownian CRT.

So far, we have concentrated on a particular class of (non-unitary) multicritical points.
We know however in the language of 2DQG how to reach unitary multicritical points corresponding
to unitary conformal field theories coupled to 2DQG [\xref\DOU,\xref\CRI]. Examples are the critical point 
of the Ising model on large random maps \BMS\ or multicritical points of hard objects with
generalized exclusion rules on large bipartite random maps \HARGRA. All these models happen
to have reformulations in terms of decorated discrete trees and one could hope that
a limiting process could produce continuous random trees in a new universality class,
hopefully now with a positive measure. 

Finally, we have limited our study to unlabeled trees but a generalization 
to embedded MCRT's, i.e. multicritical equivalents of the Brownian Snake, also seems 
reachable by the same techniques.
{}From the exact results of ref.\GEOD\ for discrete maps labeled by the geodesic distance
to an origin, we expect that the associated tree model should have both the underlying tree
and the labels simultaneously multicritical. This may be seen as follows: the fractal dimension
$2k$ of these maps relates the size $N$ of the map to the typical value $D$ of the geodesic distance
between vertices in the map via $N\sim D^{2k}$ \GEOD. In the tree language, $N$ is still the size of
the tree, now related to the typical value $L$ of the generation via $N\sim L^{d_k}$, while
$D$ is the typical value of the labels. We deduce that $L\sim  D^{2k/d_k}=D^{2(k-1)}$, hence the
scaling of labels with generation displays a non trivial fractal dimension $2(k-1)$
instead of $2$ for usual diffusion.  In other words, the embedding of the
associated MCRT should involve a multicritical diffusion process.
A similar phenomenon occurs in 2DLG when passing from the notion of distance $L$ that we used here,
defined as the natural distance on the associated tree (see. fig.\lorenty-(c)) to the notion of
``time lapse" $t$, namely the number of slices separating two points (see fig.\lorenty-(a)). 
As the fractal "time" dimension of $k$-multicritical 2DLG is $k$ \LORGRA, we necessarily have 
the anomalous scaling $L\sim t^{k-1}$.

\bigskip
\noindent{\bf Acknowledgments:} 
We thank J.-M. Luck for helpful discussions. 
All the authors acknowledge support from the Geocomp project (ACI Masse de donn\'ees) and
from the ENRAGE European network, MRTN-CT-2004-5616. J.B. acknowledges financial support 
from the Dutch Foundation for Fundamental Research on Matter (FOM). P.D.F. acknowledges support
from the ENIGMA European network, MRTN-CT-2004-5652 and from the ANR program GIMP, 
ANR-05-BLAN-0029-01.
\listrefs
\end